\documentclass[11pt,a4paper]{article}

\usepackage[utf8]{inputenc}
\usepackage{fullpage}

\usepackage[english]{babel}

\usepackage{amsmath, amsfonts, amssymb}
\numberwithin{equation}{section}

\usepackage{cite}
\usepackage{xspace}

\usepackage{hyperref}

\usepackage{tensind}
\tensordelimiter{;}
\tensorformat{s}
\begingroup\lccode`~=``
\lowercase{\endgroup\def~}#1 {;#1;}
\mathcode``="8000

\usepackage{cleveref}

\DeclareFontFamily{OT1}{pzc}{}
\DeclareFontShape{OT1}{pzc}{m}{it}%
{<-> s * [1.15] pzcmi8t}{}
\DeclareMathAlphabet{\mathpzc}{OT1}{pzc}{m}{it}

\makeatletter
\g@addto@macro\bfseries{\boldmath}
\makeatother

\newenvironment{formula}%
{\begin{equation}\begin{aligned}}%
{\end{aligned}\end{equation}\ignorespacesafterend}

\renewcommand{\bar}[1]{\overline{#1}}

\DeclareMathOperator{\re}{Re}
\DeclareMathOperator{\im}{Im}
\renewcommand{\Re}{\re}
\renewcommand{\Im}{\im}

\newcommand{\ii}{\mathrm{i}}

\newcommand{\cC}{{\mathcal{C}}}
\newcommand{\cD}{{\mathcal{D}}}

\newcommand{\cG}{{\mathcal{G}}}

\newcommand{\cL}{{\mathcal{L}}}
\newcommand{\cM}{{\mathcal{M}}}
\newcommand{\cN}{{\mathcal{N}}}

\newcommand{\cS}{{\mathcal{S}}}

\newcommand{\cV}{{\mathcal{V}}}

\newcommand{\pP}{{\mathpzc{P}}}

\newcommand{\bbC}{{\mathbb{C}}}
\newcommand{\bbD}{{\mathbb{D}}}

\newcommand{\bbG}{{\mathbb{G}}}

\newcommand{\bbM}{{\mathbb{M}}}
\newcommand{\bbN}{{\mathbb{N}}}

\newcommand{\bbR}{{\mathbb{R}}}
\newcommand{\bbS}{{\mathbb{S}}}

\newcommand{\bbZ}{{\mathbb{Z}}}

\newcommand{\CP}{\mathbb{CP}}

\newcommand{\Eseven}{\ensuremath{\mathrm{E}_{7(7)}}\xspace}

\newcommand{\SPEM}{\ensuremath{{\mathrm{Sp}(2n_{\rm v},\mathbb{R})}}\xspace}
\newcommand{\ESU}{\ensuremath{\Eseven/\SU(8)}\xspace}

\newcommand{\Gg}{\ensuremath{\mathrm{G}_{\rm gauge}}\xspace}

\newcommand{\fS}{\ensuremath{\mathfrak{S}}\xspace}

\def\SL(#1){\ensuremath{\mathrm{SL}(#1)}}
\def\PSL(#1){\ensuremath{\mathrm{PSL}(#1)}}
\def\GL(#1){\ensuremath{\mathrm{GL}(#1)}}
\def\Sp(#1){\ensuremath{\mathrm{Sp}(#1)}}
\def\USp(#1){\ensuremath{\mathrm{USp}(#1)}}
\def\SU(#1){\ensuremath{\mathrm{SU}(#1)}}
\def\PSU(#1){\ensuremath{\mathrm{PSU}(#1)}}
\def\SUs(#1){\ensuremath{\mathrm{SU}^*\kern-1pt(#1)}}
\def\U(#1){\ensuremath{\mathrm{U}(#1)}}
\def\SO(#1){\ensuremath{\mathrm{SO}(#1)}}
\def\PSO(#1){\ensuremath{\mathrm{PSO}(#1)}}
\def\SOs(#1){\ensuremath{\mathrm{SO}^*\kern-1pt(#1)}}
\def\SOp(#1){\ensuremath{\mathrm{SO}^+\kern-1pt(#1)}}
\def\O(#1){\ensuremath{\mathrm{O}(#1)}}
\def\Op(#1){\ensuremath{\mathrm{O}^+\kern-1pt(#1)}}
\def\CSO(#1){\ensuremath{\mathrm{CSO}(#1)}}
\def\CSOs(#1){\ensuremath{\mathrm{CSO}^*\kern-1pt(#1)}}
\def\ISO(#1){\ensuremath{\mathrm{ISO}(#1)}}

\def\fsl(#1){\ensuremath{\mathfrak{sl}(#1)}}
\def\su(#1){\ensuremath{\mathfrak{su}(#1)}}
\def\sus(#1){\ensuremath{\mathfrak{su}^*\!(#1)}}
\def\sp(#1){\ensuremath{\mathfrak{sp}(#1)}}
\def\so(#1){\ensuremath{\mathfrak{so}(#1)}}
\def\sos(#1){\ensuremath{\mathfrak{so}^*\!(#1)}}
\def\cso(#1){\ensuremath{\mathfrak{cso}(#1)}}
\def\csos(#1){\ensuremath{\mathfrak{cso}^*\!(#1)}}
\def\iso(#1){\ensuremath{\mathfrak{iso}(#1)}}

\def\GX{\ensuremath{{\mathrm{G}_X}}\xspace}
\def\Hm{\ensuremath{{\mathrm{H_m}}}\xspace}

\def\Gd{\ensuremath{\cG_{\rm d}}\xspace}
\def\Gm{\ensuremath{{\cG_{\rm m}}}\xspace}

\newcommand{\nv}{\ensuremath{n_{\rm v}}}

\begin{document}

\begin{titlepage}
\thispagestyle{empty}

\begin{flushright}
\hfill{Nikhef-2015-{047}}
\end{flushright}
	
\vspace{2cm}
	
\begin{center}
{\huge{\bf\boldmath Electric--magnetic deformations\\[1ex] of $\rm D=4$ gauged supergravities}}

\vspace{40pt}
{\bf Gianluca~Inverso}\footnotemark

\vspace{25pt}
{\it Nikhef, Science Park 105, 1098 XG Amsterdam, The Netherlands}

\vspace{15ex}
ABSTRACT
\end{center}

We discuss duality orbits and symplectic deformations of $D=4$ gauged supergravity theories, with focus on $\cN\ge2$.
We provide a general constructive framework for computing symplectic deformations starting from a reference gauging, and apply it to many interesting examples.
We prove that no continuous deformations are allowed for Fayet--Iliopoulos gaugings of the $\cN=2$ STU model and in particular that any $\omega$ deformation is classically trivial.
We further show that although in the $\cN=6$ truncation of $\SO(8)$ maximal supergravity the $\omega$ parameter can be dualized away, in the `twin' $\cN=2$ truncation $\omega$ is preserved and a second, new deformation appears.
We further provide a full classification and appropriate duality orbits of certain $\cN=4$ gauged supergravities, including all inequivalent $\SO(4)^2$ gaugings and several non-compact forms.

\vspace{10pt}

\footnotetext{email: \texttt{g.inverso@nikhef.nl}}
\end{titlepage}


\section{Introduction}

Supergravity theories play an essential role in our understanding of string theory, providing not only effective actions in appropriate regimes, but also many insights into its consistent backgrounds, as well as its symmetries and dualities.
An important aspect of their study is the classification of their gauged deformations:
certain gauged supergravities can arise as consistent truncations of the ten- and eleven-dimensional theories, and
not only can they be exploited to generate solutions to the equations of motion of the latter, but they also offer many insights on the exceptional symmetries and hidden structures that we expect to be present in string and M-theory.
For certain theories with a high amount of supersymmetry, a larger set of gaugings appear to be associated with so-called non-geometric compactifications, namely backgrounds that are not described by any conventional formulation of supergravity, but can be argued to exist based on duality arguments.
How to describe these non-geometric settings in a satisfying way is generally still an open question, and a better understanding of gauged supergravity should serve as guidance and as a challenge for such investigations.

Certainly one essential step in a better understanding of these models is to develop systematic methods to study and classify inequivalent gaugings of supergravity theories.
Thanks to the embedding tensor formalism \cite{Nicolai:2000sc,deWit:2002vt,deWit:2005ub}, such a task can be reduced to an algebraic and group-theoretical problem, though the explicit construction of all inequivalent gaugings of a given theory remains highly non-trivial.
A non-vanishing embedding tensor breaks explicitly the group $\cG$ of global symmetries of a theory to a subgroup \Gg that becomes local.
The classification of all inequivalent gaugings then requires one to find all the consistent embedding tensors and organize them in equivalence classes under the action of the broken $\cG$.
When $\cG$ is related to duality groups inherited from string theory, these equivalence classes are referred to as duality orbits.
This classification of $\cG$- or duality orbits has been carried out completely only for certain maximal and half-maximal supergravities in high dimensions \cite{FernandezMelgarejo:2011wx,Dibitetto:2012rk}, while
in lower dimensions the number of free parameters and the large dimension of $\cG$ for the most interesting cases make such a computation hard to carry out directly and exhaustively.

Despite the absence of full explicit classifications, some of the most striking surprises in the recent studies of gauged supergravities come from four-dimensional theories.
In \cite{DallAgata:2012bb}, it was proven that the \SO(8) gauged maximal supergravity of de Wit and Nicolai \cite{deWit:1981eq,deWit:1982ig} is only one element of an infinite family of models sharing the same gauge group but differing in the electric-magnetic embedding of its gauge-connection.
These so-called ${\omega}$-deformed \SO(8) theories all exhibit a fully supersymmetric and \SO(8)-invariant anti-de Sitter vacuum, just like the original model, and also share the same mass spectra at such extremum.
Other extrema however differ, because the higher order couplings depend non-trivially on the deformation parameter ${\omega}$.
Many further studies have unveiled the physical differences between the deformed models \cite{Borghese:2012qm,Borghese:2012zs,Borghese:2013dja,deWit:2013ija,Guarino:2013gsa,Tarrio:2013qga,Anabalon:2013eaa,Gallerati:2014xra,Cremonini:2014gia,Borghese:2014gfa,Wu:2015ska,Pang:2015mra}.
Since the original \SO(8) model can be uplifted to eleven-dimensional supergravity compactified on a seven-sphere, it is natural to ask whether the new theories also admit an uplift.
In \cite{deWit:2013ija} and \cite{Lee:2015xga} it was shown that there is no uplift to the same geometric background as the original \SO(8) model.
Thus, if an M-theory embedding of ${\omega}$ exists, it must be intrinsically non-geometric.

Deformations similar to the one of \SO(8) were soon identified and studied for other gauged maximal supergravities \cite{DallAgata:2012sx,Catino:2013ppa,DallAgata:2014ita}, giving rise among other things to the first slow-roll solutions in maximal supergravity and to a vast landscape of Minkowski models with spontaneously broken supersymmetry.
An important result was the proof of the existence of a discrete deformation for the \ISO(7) gauging \cite{DallAgata:2011aa,DallAgata:2014ita}.
The string theory origin and CFT dual of such models were recently identified in \cite{Guarino:2015jca,Guarino:2015qaa,Guarino:2015vca}, were it was shown that the deformation is associated with Romans mass in IIA supergravity compactified on a six-sphere.\footnote{This gauged supergravity is often referred to as `dyonic \ISO(7) gauged supergravity'. It should be pointed out that this terminology is misleading, because all gauging charges are mutually local by construction.}

A crucial aspect in the classification of duality orbits of gaugings is the explicit construction of those identifications that reduce the range of (or eliminate entirely) the deformation parameters.
Such transformations cannot be identified nor proven to exist by arguments based only on duality invariants, yet knowing their form can be crucial.%
\footnote{For instance, when defined in an electric frame the range of ${\omega}$ for the \SO(8) theory is reduced by parity and by a certain \Eseven transformation \cite{DallAgata:2011aa} of the scalar fields inducing the $\bbZ_2$ outer automorphism of the gauge group, combined with a redefinition of the vector fields by the same automorphism \cite{DallAgata:2012bb,deWit:2013ija,DallAgata:2014ita}.
Employing these local field redefinitions, the ${\omega}={\pi}/4$ \SO(8) gauged supergravity can also be lifted to eleven dimensions.
In particular, self duality and anti self duality of certain \SO(8) four-forms are reversed by the outer automorphism, which is necessary for the proof of the `Clifford property' of \cite{deWit:2013ija} when ${\omega}=\pi/4$.
This clarifies a mismatch between the values of ${\omega}$ that are described as liftable in \cite{deWit:2013ija} and \cite{Lee:2015xga}, and the range of inequivalent theories in four dimensions.}
A systematic, constructive approach to the classification of duality orbits of gaugings of four-dimensional maximal supergravity was introduced in \cite{DallAgata:2014ita}, which defined the concept of a deformation of the symplectic frame of a four-dimensional theory, compatible with a certain fixed choice of embedding tensor (for short, \emph{symplectic deformation}).
This space parameterizes inequivalent gaugings of a theory, all sharing the same gauge group but differing in the specifics of their couplings.
All field redefinitions that give rise to identifications between models are also encoded explicitly in the space of symplectic deformations.
Equivalent to the space of duality orbits of gaugings of a certain group \Gg is a certain double coset-space called the `reduced' $\mathfrak S$-space
\begin{formula}
\mathfrak S_{\rm red} = \cS_{\Sp(56,\bbR)}(`[c]X_MN^P )\ \backslash\ \cN_{\Sp(56,\bbR)}(\Gg)\ /\ \cN_{\Eseven\rtimes\bbZ_2}(\Gg)\ , 
\end{formula}
where $`X_MN^P $ is the embedding tensor, \Sp(56,\bbR) is the group of Gaillard--Zumino duality redefinitions of the vectors of maximal supergravity, and $\cS, \cN$ indicate stabilizers and normalizers respectively.
When the gaugings are defined in an electric frame, it can be meaningful to only regard as equivalent those theories that are related by local field redefinitions.
In this case a full space $\mathfrak S$ is defined, where in the left quotient only \GL(28,\bbR) local redefinitions of the physical vector fields are allowed.
As a simple example of the difference between the two, $\mathfrak S$ treats as inequivalent theories also those that differ by a shift in a (constant) theta-angle, while $\mathfrak S_{\rm red}$ captures all and only the deformations that affect the classical equations of motion.

Surprises similar to those encountered in maximal supergravity surely hide in the gaugings of less supersymmetric theories.
In this paper we extend the approach of \cite{DallAgata:2014ita} to non-maximal gauged supergravities in four dimensions, providing classifications of large sets of inequivalent models.
We will give a general discussion followed by many physically relevant examples.
Some of the latter are strictly related to consistent truncations of the ${\omega}$-deformed \SO(8) gauged maximal supergravity and clarify what couplings are really non-trivially affected by ${\omega}$ and render non-viable an uplift to the standard geometric eleven-dimensional supergravity.
We will show that ${\omega}$ is trivialized in certain truncations where it was expected to be relevant, and survives in others.
Many further examples are provided, not related to the \SO(8) theories, with particular attention to $\cN=4$ supergravities.

We will mostly focus on the case where the global symmetries $\cG$ of the theory under consideration factor into electric-magnetic duality symmetries and other `matter' symmetries, which is sufficient to encompass extended supersymmetric models.
As we will explain in the following sections, when a gauging of matter symmetries is involved, the concepts of symplectic deformations and of $\cG$-orbits differ slightly, although both allow in principle a full classification of gaugings of a theory.
The computation of symplectic deformations is favored because of its group-theoretical nature, where the quadratic constraints of the embedding tensor only need to be solved once for each choice of gauge group.

This paper is organized as follows.
In section~\ref{sec2} we briefly review the embedding tensor formalism, discuss its $\cG$-orbits and define the space of symplectic deformations for $\cN\ge2$ gauged theories.
In section~\ref{sec3} we discuss the truncation of the \SO(8) gauged maximal supergravities to $\cN=6$ supergravity and its $\cN=2$ sibling sharing the same bosonic content.
We make a surprising discovery in that we find that ${\omega}$ is trivial in the former but not in the latter, where it is joined by a second deformation.
Section~\ref{sec4} is dedicated to several examples of gauged half-maximal supergravities, and section~\ref{sec5} describes all gaugings of the STU-model, showing in particular that no non-trivial ${\omega}$-deformation is present.
We conclude in section~\ref{sec6}.

\section{Symplectic deformations with gauged matter symmetries}\label{sec2}

\subsection{General gaugings and $\cG$-orbits}

The embedding tensor formalism for $D=4$ gauged theories was formulated in \cite{deWit:2005ub} and soon applied to the construction of the maximal \cite{deWit:2007mt} and half-maximal \cite{SchonWeidner} gauged supergravities.
The formalism was also applied to $\cN=2$ rigid theories \cite{deVroome:2007zd} and supergravity in the superconformal formulation \cite{deWit:2011gk}.

Given the group $\cG$ of global symmetries of a theory, 
we use the set of electric and magnetic vector potentials $A_{\mu}^M$ ($M=1,\ldots,2\nv$, where $\nv$ is the number of physical vectors) to gauge a subgroup $\Gg\subseteq\cG$. 
The choice of gauging is completely encoded in an embedding tensor $`{\Theta}_M^{\alpha} $, such that covariant derivatives take the schematic form $\cD_{\mu} \equiv {\partial}_{\mu} - A_{\mu}^M`{\Theta}_M^{\alpha} {\tau}_{\alpha} $, where ${\tau}_{\alpha} $ are $\cG$ generators in any appropriate representation.
Notice that ${\alpha}$ is an index in the adjoint of $\cG$, while $M$ enumerates the vectors and therefore also forms a representation of \SPEM, the group of electric-magnetic duality transformations \cite{Gaillard:1981rj}.
The subgroup $\Gd\subseteq\cG$ that is non-trivially represented on the vectors is embedded in \SPEM, and corresponds to duality symmetries of the theory.
It also acts on the other fields by inducing isometry transformations of the scalar manifold.

All modifications to the couplings of a (supergravity) theory that are due to a gauging can be expressed in terms of $`{\Theta}_M^{\alpha} $.
The embedding tensor satisfies certain linear and quadratic constraints that guarantee consistency of the resulting gauged theory:
\begin{formula}
\label{general emb tens constraints}
`{\Omega}^MN `{\Theta}_M^{\alpha} `{\Theta}_N^{\beta} &= 0\qquad\text{(locality)}, \\
`{\Theta}_M^{\alpha} `{\Theta}_N^{\beta} `f_{\alpha}{\beta}^{\gamma} + `{\Theta}_M^{\alpha} `t_{\alpha}N^P `{\Theta}_P^{\gamma} &= 0\qquad\text{(closure)}, \\
`{\Theta}_P^{\alpha} `t_{\alpha}_M^P = `{\Theta}_(M^{\alpha} `t_{\alpha}N^Q `{\Omega}_P)Q &= 0\qquad\text{(susy/counting of d.o.f.)},
\end{formula}
where ${\Omega}_{MN}$ is the symplectic invariant and $`t_{\alpha}M^N $ are the generators of $\Gd$ in the symplectic representation that acts on the vector fields.
Notice that this need not be a faithful representation of the full \Gg.
We denote \GX the subgroup that is faithfully represented on vectors.
Defining $`X_MN^P \equiv `{\Theta}_M^{\alpha} `t_{\alpha}N^P $, the above constraints imply
\begin{formula}
\label{X constraints}
`{\Omega}^MN `X_MP^Q `X_NR^S = 0 , \\
[X_M,\ X_N]`{}_P^Q +`X_MN^R `X_RP^Q = 0 , \\
`X_PM^P = `X_(MNP) = 0 .
\end{formula}
If $\Gg=\GX$, i.e. only duality symmetries are gauged, $`X_MN^P $ entirely defines the gauging and the above constraints are necessary and sufficient for consistency.
Otherwise, $`{\Theta}_M^{\alpha} $ contains extra non-vanishing entries associated with the gauging of non-duality symmetries and \eqref{general emb tens constraints} imposes further consistency constraints.
In any case, the gauging of \GX is by definition completely encoded in a tensor $`X_MN^P $.
It is also important to stress that the first two constraints on $`X_MN^P $ are equivalent when the linear one is satisfied.

Let us now discuss how all gaugings of a chosen \GX can be characterized.
We can decompose $`X_MN^P $ in terms of generators $`t_xM^P $, $x$ being an adjoint index for \GX, and a `small' embedding tensor $`{\vartheta}_M^x $ such that 
\begin{formula}
`X_MN^P =`{\vartheta}_M^x `t_xN^P \ .
\end{formula}
Clearly, $`{\vartheta}_M^x $ specifies the choice of gauge connection for \GX.
The consistency constraints \eqref{X constraints} then reduce to linear equations in $`{\vartheta}_M^x $:
\begin{align}
\label{vector constraints}
`{\vartheta}_M^x `f_xy^z  = `t_yM^N `{\vartheta}_N^z \ ,\qquad
`{\vartheta}_M^x `t_xN^M  = `{\vartheta}_(M^x `t_xNP) = 0\ ,
\end{align}
where $`f_xy^z $ are the structure constants of \GX.
Given one solution of these constraints, every other one is obtained by the action on $`X_MN^P $ of symplectic transformations in
\begin{formula}
\label{symp defo vector only}
\cS_{\SPEM}(`[c]X_MN^P )\ \backslash\ \cN_\SPEM(\GX)\ ,
\end{formula}
where $\cS_{\rm G}(X)$ is the stabilizer of X in the group G, and $\cN_{\rm G}(K)$ is the normalizer in a group G of its subgroup K.%
\footnote{Contrary to \cite{DallAgata:2014ita}, we never include overall rescalings of the embedding tensor in $\cS_{\SPEM}(`[c]X_MN^P )$.}
This result is obtained following the same procedure as in $\cN=8$ supergravity \cite{DallAgata:2014ita}.
In an appropriate electric frame, one solution of the constraints can be taken to be $`{\vartheta}0_M^x =`{\delta}_M^x $.

We can try to treat the gauging of the full $\Gg\subseteq\cG$ in a similar way.
Assuming that the choice of gauge group has been made, we introduce its generators ${\tau}_A$, $A=1,\ldots,\textrm{dim}\Gg$, structure constants $`f_AB^C $, and a small embedding tensor $`{\vartheta}_M^A $.
This time, however, we must solve the more general consistency constraints \eqref{general emb tens constraints}.
A first difference is that these result in a set of quadratic equations for $`{\vartheta}_M^A $, rather than linear ones.
A more crucial difference with the previous case (and with maximal supergravity in particular) is that here is no analogue of \eqref{symp defo vector only} that can characterize all consistent $`{\vartheta}_M^A $ group theoretically.
Intuitively, the problem is that the full \Gg is not embedded in \SPEM.
There is therefore no equivalent of $`X_MN^P $ that entirely defines the gauging, on which symplectic transformations can act to induce a change of the \emph{full} $`{\vartheta}_M^A $.
If we nevertheless assume to have solved the linear and quadratic constraints on $`{\vartheta}_M^A $ for a given \Gg, we can ask which of the resulting solutions give rise to inequivalent theories.
To answer this question, we should quotient out any $\cG$ transformation that maps \Gg to itself.
Namely, we should compute the action of the normalizer $\cN_\cG(\Gg)$ on the general solutions $`{\vartheta}_M^A $, thus obtaining the classification of $\cG$-orbits of gaugings of \Gg.
This is the obvious generalization of the computation of duality orbits of maximal and half-maximal gauged supergravities, when the global symmetries of the theory are larger than the duality symmetries \Gd.

Given that the space of consistent gauge connections has no group-theoretical description, the characterization of the $\cG$-orbits can only be carried out on a case-by-case basis, computing explicitly the action of  $\cG$ transformations on the embedding tensor.
What we will do instead is to pose an analogous but subtly different problem, the answer to which can be phrased in terms of a certain subset of symplectic transformations that can be computed explicitly.

\subsection{Symplectic deformations of $\cN\ge2$ gauged theories}

From now on we will focus on theories where the global symmetries decompose in a direct product of duality and `matter' symmetries, the latter leaving the vector fields invariant:\footnote{The non-minimal couplings between scalars and vectors can be parameterized in terms of a symmetric matrix $\cM({\phi})_{MN}$ and beyond those symplectic transformations that induce isometries on the scalar manifold, there can be extra symmetries in \Gd associated with $U\in\U(\nv)\subset\SPEM$ if $[U,\cM({\phi})]=0\ \forall {\phi}$.}
\begin{formula}
\cG = \Gd \times \Gm.
\end{formula}
In particular, this situation arises in theories with extended supersymmetry.
It is convenient for our purposes to first consider the consistent gauging of some group $\GX\subset\Gd$, entirely specified by $`X_MN^P $, and then the further coupling of matter symmetries $\Hm\subset\Gm$ to the vector fields, so that the final gauging is a maximal subgroup $\Gg\subseteq\GX\times\Hm$.
Assuming $\GX$ and $\Hm$ have been chosen and fixed, we can rewrite the constraints \eqref{general emb tens constraints} in terms of $`X_MN^P $ and a small embedding tensor $`{\theta}_M^a $, with $a$ running along the adjoint of $\Hm$.
Beyond the constraints \eqref{X constraints} for $`X_MN^P $, we now also have
\begin{formula}
\label{contraints matter emb tens}
`{\theta}_M^a `{\theta}_M^b `f_ab^c = -`X_MN^P `{\theta}_P^c ,\qquad
`{\theta}_M^a `{\theta}_N^b `{\Omega}^MN = `{\theta}_M^a `X_NP^Q `{\Omega}^MN =0.
\end{formula}
One advantage of this decomposition is that the consistency constraints on  $`X_MN^P $ are unchanged.
Finding all solutions $`{\theta}_M^a $ that satisfy the quadratic constraints for a given $`X_MN^P $ can be difficult, although there are some obvious simplifications.
For instance, the non-Abelian part of \Hm must be gauged by vector fields that already gauge an isomorphic subgroup in \GX.
This means that the above constraints only need to be solved explicitly for central extensions and Abelian factors in \Hm.\footnote{As will be clear in the following discussion, the rank of $`{\theta}_M^a $ is fixed and maximal.}
Notice however that different choices of $`{\theta}_M^a $ compatible with the same $`X_MN^P $ might modify the embedding of \Gg into $\GX\times\Hm$, possibly mapping it to an inequivalent one.
Hence, the classification of solutions of \eqref{contraints matter emb tens} differs from that of $\cG$-orbits of gauge connections for a fixed gauge group.
This distinction is only relevant when matter symmetries are gauged.

We are now ready to define our problem.
The question we ask is what set of symplectic frames are compatible with the introduction of the \emph{same} tensors $`X0_MN^P $, $`{\theta}0_M^a $.
The introduction of a fixed embedding tensor in different symplectic frames will in general affect the resulting equations of motion, thus yielding inequivalent gauged theories.
Let us rephrase this problem from a perspective similar to the discussion of inequivalent gauge-connections we have carried out so far.
Any two frames are related by a symplectic transformation $`N_M^P $.
Instead of explicitly classifying all Lagrangians associated with the action of these transformations on the kinetic terms and moment-couplings of the vector fields, we can decide to always revert to a fixed choice of symplectic frame \emph{after} $`X0_MN^P $ and $`{\theta}0_M^a $ have been turned on, by acting with the inverse transformation $`N^-1_M^P $ on both the vector couplings and the embedding tensor.
As a result, computing symplectic deformations is equivalent to classifying all transformations $`N_M^P $ such that
\begin{formula}
\label{symp defo action on emb tens}
`X0_MN^P \to `N_M^Q `N_N^R `X0_QR^S `N-1_S^P ,\quad
`{\theta}0_M^a \to `N_M^N `{\theta}0_N^a ,
\end{formula}
yield a consistent gauging in a fixed symplectic frame.
We now notice that the most general $`X_MN^P $ gauging \GX is obtained from a reference one by transformations $`N_M^N \in \cN_\SPEM(\GX)$, and that in this case also \eqref{contraints matter emb tens} is satisfied.
We conclude that symplectic deformations are specified by elements of $\cN_\SPEM(\GX)$, in full analogy with $\cN=8$ supergravity.
In contrast to maximal supergravity, the resulting models are allowed to span inequivalent embeddings of \Gg in $\GX\times\Hm$, if more than one exists.
On the other hand, this is not an exhaustive classification of $\cG$-orbits of gaugings of a fixed \Gg discussed in the previous section, making the two concepts inequivalent but complementary.
The advantage of symplectic deformations is that we have a general framework to construct them.

\subsection{The quotients}

The set of gaugings we are interested in is determined by $\cN_\SPEM(\GX)$ even when we have a gauging of matter symmetries.
The presence of the latter however affects what elements of $\cN_\SPEM(\GX)$ should be regarded as giving rise to equivalent theories.

Let us keep reasoning in terms of the fixed-frame approach, in which symplectic deformations act on the embedding tensor according to \eqref{symp defo action on emb tens}.
In absence of gauged matter symmetries we can write 
\begin{formula}
N\cong S\,N,\qquad S\in \cS_\SPEM(`X0_MN^P ),\qquad
N\in\cN_\SPEM(\GX),
\end{formula}
since the transformation $S$ has by definition no effect on the embedding tensor.
The most natural generalization of this identification would be to require
\begin{formula}
N\cong S\,N,\qquad S\in \cS_\SPEM(`X0_MN^P ,`{\theta}0_M^a ),
\end{formula}
as proposed in \cite{DallAgata:2014ita}.
We can actually quotient out a larger group of transformations.
The key point is that the constraints \eqref{contraints matter emb tens} are preserved under any transformation such that
\begin{formula}
`S_M^N `{\theta}0_N^a = `{\theta}0_M^b `m_b^a ,\qquad 
S\in\cS_\SPEM(`X0_MN^P ),\qquad
`m_a^b \in \frac{\cN_\Gm(\Hm)}{\cC_\Gm(\Hm)}.
\end{formula}
The latter quotient is isomorphic to a subgroup of $\mathrm{Aut}(\Hm)$, hence it can be represented in the adjoint of $\Hm$ as specified.
Clearly, any $`m_a^b \in \mathrm{Aut}(\Hm)$ would preserve the embedding tensor constraints, since by definition $`m_a^d `m_b^e `f_de^f m^{-1}`{}_f^c = `f_ab^c $.
However, in the Lagrangian such automorphisms must be induced by a field redefinition of the matter fields obtained from the action of the (broken) \Gm global symmetries.
One way to see this is to look at the covariant derivative for some matter fields ${\phi}$:
\begin{formula}
\cD_{\mu} {\phi}  \equiv {\partial}_{\mu} {\phi} -A_{\mu}^M `{\theta}0_M^a {\tau}_a({\phi})  \ \  \to\ \
{\partial}_{\mu}  {\phi} -A_{\mu}^M `{\theta}0_M^b `m_b^a {\tau}_a({\phi}) ,
\end{formula}
where ${\phi}$ is some matter field and ${\tau}_a({\phi})$ are the infinitesimal variations of the fields under \Hm.
Clearly the left and right hand side can only be equivalent if $`m_a^b $ can be reabsorbed in a redefinition of ${\phi}$ that also leaves invariant all couplings unrelated to the gauging (in particular, it must be an isometry of the scalar manifold to preserve the kinetic terms).
We thus define the subgroup of $\cN_\SPEM(\GX)$ that can be appropriately quotiented away:
\begin{formula}
\label{bbS}
\mathbb S(X^0,\, {\theta}^0) \equiv \left\{
S \in \cS_\SPEM(X^0)\ |\ 
S {\theta}^0 = {\theta}^0 m ,\ 
m \in \frac{\cN_\Gm(\Hm)}{\cC_\Gm(\Hm)}
\right\}.
\end{formula}

A second set of transformations to quotient out is associated with duality symmetries \Gd.
In full analogy with maximal supergravity, these identifications are obtained by imposing
\begin{formula}
N\cong N D, \qquad D\in \cN_{\Gd}(\GX).
\end{formula}
In this case no further changes are needed.

Let us now comment on the point of view in which a fixed choice of $`X0_MN^P $ and $`{\theta}0_M^a $ is made, and symplectic deformations affect the frame where these tensors are introduced.
For instance, $`X0_MN^P $ and $`{\theta}0_M^a $ could involve electric vectors only, and we would be classifying all electric frames compatible with such gauge couplings.
The quotients that we need to perform are necessarily the same (the two approaches are equivalent), but their interpretation changes.
The quotient by \Gd now corresponds to redefinitions of the scalar fields only (non-linearly realized on the fermion fields, too), rather than duality transformations.
Again, this is analogous to maximal supergravity \cite{DallAgata:2014ita}.
The elements of $\mathbb S(X^0,\, {\theta}^0)$ now have a non-trivial interpretation: they correspond to electric-magnetic redefinitions of the vector fields such that any effect on the gauge interaction terms can be removed by a redefinition of the matter fields, as described above.
Therefore, these electric-magnetic redefinitions do not affect the equations of motion and can be safely quotiented out as we have done.
However, some of these redefinitions are non-local, and in an electric frame we might want to only allow for \emph{local} field redefinitions instead.
If this is the case, \eqref{bbS} must be substituted with its subgroup of \GL(\nv,\bbR) redefinitions of the electric vectors:
\begin{formula}
\label{bbG}
\mathbb G(X^0,\, {\theta}^0) \equiv \left\{
S \in \cS_{\GL(\nv,\bbR)}(X^0)\ |\ 
S {\theta}^0 = {\theta}^0 m ,\ 
m \in \frac{\cN_\Gm(\Hm)}{\cC_\Gm(\Hm)}
\right\}.
\end{formula}
To give a simple example, quotienting by $\mathbb G(X^0,\, {\theta}^0)$ rather than by $\mathbb S(X^0,\, {\theta}^0)$ makes us regard as inequivalent theories that differ by shifts in theta-terms.

\subsection{Parity}\label{sec:parity}

If the ungauged theory we start with admits a parity symmetry, it must have an action on the scalar manifold and hence act as an automorphism of $\cG$.
Parity acts on \Gd as an anti-symplectic transformation $`P_M^N $inducing a $\bbZ_2$ automorphism \cite{Ferrara:2013zga,Aschieri.TMP,DallAgata:2014ita}.
In the matter sector, parity might induce an automorphism of \Hm, too.
Denoting $`p_a^b $ this automorphism, if it exists, we make use of the fact that the locality constraints allow us to map two matrices $`{\theta}_M^a $, $`{\theta}'_M^a $ to each other both via a symplectic and an anti-symplectic transformation.
In particular, we can define $\hat P`{}_M^N $ such that
\begin{formula}
`{\theta}0_M^b `p_b^a = \hat P`{}_M^N `{\theta}0_N^a .
\end{formula}

Combining these observations, parity induces an extra identification on $\cN_\SPEM(\GX)$ only when the following conditions are satisfied:
\begin{enumerate}
\item the ungauged theory has a parity symmetry and there is a representative $`P_M^N $ that normalizes \GX,
\item the action of parity on \Gm can be taken to induce a transformation $`p_a^b \in \mathrm{Aut}(\Hm)$ (i.e. to normalize \Hm),
\item the induced $\hat P$ transformation can be chosen to stabilize $`X0_MN^P $.
\end{enumerate}
In this case, we have the extra identification
\begin{formula}
N\cong \hat P N P\ ,\qquad N\in \cN_\SPEM(\GX)\ .
\end{formula}
As a byproduct, a given gauging defined by $`X0_MN^P ,\ `{\theta}0_M^a $ admits a parity symmetry only if 
$
P\hat P \in \mathbb S(`X0 ,\, `{\theta}0 ).
$
When no matter symmetries are gauged, these requirements reduce to $`P_M^N $ normalizing \GX, because in that case its action on $`X0_MN^P $ must be equivalent to that of an element $Q^{-1}$ of $\cN_\SPEM(\GX)$, which classifies exhaustively all gaugings of \GX.
Then, one can just define $\hat P \equiv P Q$, as in the maximal case \cite{DallAgata:2014ita}.

We have arrived at the definition of the so-called `reduced $\fS$-space':
first we define
\begin{formula}
\fS^0_{\rm red} \equiv
\mathbb S(`X0 ,\, `{\theta}0 )\ \backslash\ \cN_\SPEM(\GX)\ /\ \cN_{\Gd}(\GX),
\end{formula}
then, depending on whether the conditions above are satisfied, we have
\begin{formula}
\label{Sred general}
\fS_{\rm red} \equiv \left\{
\begin{array}{ll}
\fS^0_{\rm red}/\bbZ^\pP_2 & \text{$\exists$ parity identification},\\
\fS^0_{\rm red}        & \text{otherwise}.\\
\end{array}
\right.
\end{formula}
Finally, when we regard the symplectic deformations as choices of symplectic frames of the ungauged theory, where it is allowed to introduce a fixed choice of embedding tensor $(`X0 ,\, `{\theta}0 )$, we may want to regard as inequivalent also all those symplectic frames related by elements  $\mathbb S(`X0 ,\, `{\theta}0 )$ which, however, induce non-local redefinitions of the vector potentials,
In doing so we define a `full' $\fS$-space where the left quotient in the above formulas must then be substituted with \eqref{bbG}, and parity identifications are required to not mix the electric vector fields with their duals.

\section{Twin supergravities and the fate of the ${\omega}$-deformation}\label{sec3}

Let us now apply our framework to some interesting cases.
Maximal supergravity can be truncated to $\cN=6$ supergravity, breaking the \SU(8) local symmetry to $\SU(6)\times\SU(2)\times\U(1)$ and keeping only the $\SU(2)$-singlets \cite{Andrianopoli:2008ea}.
In the process the scalar manifold is reduced to $\SOs(12)/\U(6)$.
Interestingly, another consistent truncation of the maximal theory is achieved if the preserved bosonic field content is the same as for $\cN=6$, but now it is the \SU(6) fermionic singlets that are kept in the spectrum.
The resulting model is $\cN=2$ supergravity coupled to fifteen vector multiples, parameterizing $\SOs(12)/\U(6)$ as special K\"ahler target space \cite{Andrianopoli:2008ea,Roest:2009sn}.

The $\cN=6$ \SO(6) gauged model that arises from truncation of \SO(8) gauged maximal supergravity can be regarded as a consistent truncation of type IIA supergravity on $\CP^3$.
Correspondingly, when truncating 11d supergravity on $S^7$ to type IIA by reducing on the $S^1$ Hopf fiber of $S^1 \hookrightarrow S^7\to \CP^3$, two supersymmetries end up hidden in non-perturbative states of the IIA theory.
This is an example of `superymmetry without supersymmetry' \cite{Duff:1997qz}.
The $\cN=2,\ D=4$ supergravity twin of the $\cN=6$ \SO(6) model captures exactly the two hidden supersymmetries.
As pointed out in \cite{Andrianopoli:2008ea}, all gaugings of $\cN=6$ supergravity are obtained as truncations of the maximal theory.

It is clearly extremely interesting to study how the ${\omega}$-deformation of \SO(8) gauged $\cN=8$ supergravity is realized on these consistent truncations.
In \cite{Borghese:2014gfa} an initial analysis of the $\cN=6$ model was carried out. 
Here we complete this study, and perform the same computation for the twin $\cN=2$ model, where we find some surprises.
In particular, we will prove that the \SO(6) gauged $\cN=6$ supergravity does not admit any classically non-trivial deformations (in our language, $\fS$ is trivial) and the ${\omega}$ parameter inherited from the maximal theory is one of three parameters that can be eliminated by a change of symplectic frame, at most affecting only boundary terms and quantum corrections, thus completing the analysis of \cite{Borghese:2014gfa}.
For the twin $\cN=2$ theory instead, we will show that the ${\omega}$ parameter remains non-trivial also at the classical level, and actually a second deformation is available.

\subsection{Deformations of $\cN=6$ \SO(6) gauged supergravity}

The $\cN=6$ supergravity contains sixteen vector fields and thirty (real) scalars, transforming in the $\mathbf{15}+\mathbf{1}$ and $\mathbf{15}+\mathbf{15}$ of the \SO(6) that we gauge, respectively.
The duality symmetries of the ungauged theory are $\SOs(12)\subset\Sp(32,\bbR)$ where the vectors transform in a chiral spinorial representation.

The symplectic frame we use can be obtained starting from maximal supergravity in the \SU(8) covariant form, where one has complex linear combinations $`A_{\mu}^[IJ] $ of the vector fields and their duals, $I,J,\ldots$ being \SU(8) fundamental indices.
Similarly the scalar fields are denoted $`{\phi}^IJKL $ and give rise to the coset representative 
$ 
L({\phi}) \sim \exp\left(\begin{smallmatrix}
 & `{\phi}_IJKL \\ `{\phi}^IJKL &  
\end{smallmatrix}\right). 
$
The $\cN=6$ fields are obtained keeping only $A_{\mu}^{[ab]} $, $a,b=1,\ldots,6$, $A_{\mu}^{[78]} \equiv A_{\mu}^0 $, $`{\phi}^abcd $ and their complex conjugates.
We can then switch to a real basis for the vectors 
using a Cayley matrix
$$
\frac1{\sqrt2} \left(\begin{matrix}
{\bf1}_{16} & {\bf1}_{16} \\
-\ii {\bf1}_{16} & \ii {\bf1}_{16}
\end{matrix}\right),
$$
making only an \SO(6) subgroup of \U(6) explicitly covariant.
The scalar matrix $L(\phi)$ can be also truncated to a $32\times32$ dimensional coset representative of $\SOs(12)/\U(6)$.
The resulting electric group in this symplectic frame is $\PSL(4,\bbC)\subset\SOs(12)$.

All we need to know to compute the symplectic deformations is how \SO(6) is embedded in \Sp(32,\bbR) and the form of the embedding tensor $`X_MN^P $.
To be more precise, the fundamental of \Sp(32,\bbR) contains the adjoint of $\PSO(6)\cong\SO(6)/\bbZ_2$ embedded as:%
\footnote{Notice that fermions transform under the double cover \SO(6), rather than under its universal covering \SU(4). Their representations are indeed tensor products of the vector irrep.
This is analogous to the situation in maximal supergravity where fermions transform under $\SO(8)\underset{\rm max.}{\subset}\SU(8)$ and becomes relevant in the computation of outer automorphisms.}
\begin{formula}
t_{\PSO(6)} = \begin{pmatrix}
 `{\Lambda}_ab^cd &    &           &    \\
          & 0  &           &    \\
          &    & -`{\Lambda}_cd^ab &    \\
          &    &           & 0  \\           
 \end{pmatrix}
\end{formula}
where $ab, cd$ are now antisymmetrized pairs of vector indices of \SO(6) and $`{\Lambda}_ab^cd \in \text{adj}(\su(4))$.
We have chosen the \SO(6) invariant vector to be the last.
In the standard symplectic frame the embedding tensor takes the form (up to an overall constant)
\begin{formula}
`X_abM^N &= \cos{\omega}\,\begin{pmatrix}
 -2`{\delta}_[a^[e `{\delta}_b][c `{\delta}_d]^f] &    &               &    \\
               & 0  &               &    \\
               &    & 2`{\delta}_[a^[c `{\delta}_b][e `{\delta}_f]^d]   &    \\
               &    &               & 0  \\           
 \end{pmatrix}
,\\
`X^ab_M^N &= \sin{\omega}\,\begin{pmatrix}
 -2`{\delta}_[a^[e `{\delta}_b][c `{\delta}_d]^f] &    &               &    \\
               & 0  &               &    \\
               &    & 2`{\delta}_[a^[c `{\delta}_b][e `{\delta}_f]^d]  &    \\
               &    &               & 0  \\           
 \end{pmatrix}.
\label{N6omegaX}
\end{formula}
We have included for reference the ${\omega}$ parameter as it would appear from the truncation of the $\cN=8$ supergravity embedding tensor, but for our purposes we only need some initial choice of embedding tensor, hence from now on we fix $X^0 \equiv X|_{{\omega}=0}$.

We now compute the double quotient
\begin{formula}
\bbS(X^0)\ \backslash\ \cN_\SPEM(\PSO(6))\ /\ \cN_{\SOs(12)}(\PSO(6))).
\label{SredQuotientN6}
\end{formula}
The centralizer of \PSO(6) in \SPEM is easily computed using Schur's lemma.
It turns out to be
$
\cC_{\Sp(32,\bbR)}(\PSO(6)) = \SL(2,\bbR)\times\SL(2,\bbR)_0
$
where $\SL(2,\bbR)_0$ is the group of duality redefinitions of the ungauged vector, while the first factor acts in the same way on all the fifteen gauged vectors, thus commuting with \PSO(6), and leaves the singlet vector invariant.
The normalizer is obtained noting that $\mathrm{Out}(\PSO(6))=\bbZ_2$, and such an automorphism is indeed inherited from $\cN_{\Sp(56,\bbR)}(\PSO(8))$ in the maximal theory.
Therefore we have (we always leave the gauge group itself as understood when writing normalizers)
\begin{formula}
\cN_{\Sp(32,\bbR)}(\PSO(6)) \cong \bbZ_2 \times \SL(2,\bbR)\times\SL(2,\bbR)_0.
\end{formula}
The $\bbZ_2$ outer automorphism of an \SO(2n) group can always be realized in the vector representation as the matrix $\mathrm{diag}(-1,+1,\ldots,+1)$.
In our case it is embedded in \Sp(32,\bbR) as a diagonal matrix with $\pm1$ eigenvalues, where the negative signs are associated with the vectors $A_{\mu}^{1a},\ a=2\ldots6$ and their duals.
If we combine it with a sign flip of the ungauged vector (which is an $\SL(2,\bbR)_0$ transformation), we actually obtain an element of \U(6).
We will refer to the combined transformation as $Z\in\cN_{\U(6)}(\SO(6))$.%
\footnote{$Z$ can be mapped to the \SU(8) transformation associated with the outer automorphism of \SO(8), introduced in \cite{DallAgata:2011aa} and exploited in \cite{DallAgata:2012bb,deWit:2013ija,DallAgata:2014ita}.}

At this point a quick computation also shows that $\cN_{\SOs(12)}(\PSO(6))=\U(1)\times\bbZ_2$, the \U(1) factor being of course the center of \U(6) and $\bbZ_2$ being the $Z$ transformation we have just defined.
The \U(1) transformation is generated in \sp(32,\bbR) by
\begin{formula}
t_{\U(1)}\propto
\left(\begin{smallmatrix}
&&+\mathbf1_{15}&   \\
&&           & -3 \\
-\mathbf1_{15}&&&  \\
           &+3&&
\end{smallmatrix}\right).
\label{tU1}
\end{formula}
We also find $\bbS(X^0)=\bbZ_2\times\bbR\times\SL(2,\bbR)_0$, where the $\bbR\subset\SL(2,\bbR)$ factor corresponds to shifts in the theta-angle $\theta_{\SO(6)}$ of the gauged vectors.
Recall that we are not allowing for rescalings of the gauge coupling constant.
We must also quotient by a parity identification, which however does not affect the final result in this case.
Putting everything together, we arrive at 
\begin{formula}\label{SredN6}
(\fS_{\rm red})^{\cN=6}_{\SO(6)}\ \cong\ \bbR\ \backslash\ \SL(2,\bbR)\ /\ \U(1)\ \cong\ \bbR_+\ ,
\end{formula}
where the only remaining transformation is nothing but the rescaling of the gauge coupling constant.
We see in particular that there is no non-trivial ${\omega}$-deformation in \SO(6) $\cN=6$ gauged supergravity.
In fact, the ${\omega}$ parameter in \eqref{N6omegaX} can be set to vanish via the \U(6) transformation \eqref{tU1}.

The computation we have just made is equivalent to the classification of  duality orbits of embedding tensors gauging \SO(6), under the duality symmetry group \SOs(12).
Following the reasoning in \cite{DallAgata:2014ita}, we can instead ask what set of symplectic frames of ungauged $\cN=6$ supergravity admit the introduction of $`X0_MN^P $ as gauge couplings.
The difference is that now we regard \SOs(12) as local redefinitions of the scalar fields, and we substitute the left quotient in \eqref{SredQuotientN6} with its subgroup $\bbG(X^0)$, corresponding to \GL(16,\bbR) local field redefinitions of the physical vector potentials.
The \GL(16,\bbR) group is embedded block-diagonally in \Sp(32,\bbR).
Moreover, in this case the action of parity must be taken into account: we can borrow the anti-symplectic transformation used in maximal supergravity \cite{DallAgata:2014ita}, truncated to the $\cN=6$ fields:
\begin{formula}\label{Pmatrix}
P_M{}^N = {\hat{P}}_M{}^N = \left(\begin{matrix}
\mathbf1_{16}&\\
&-\mathbf1_{16}
\end{matrix}\right).
\end{formula}
It stabilizes $`X0_MN^P $ and induces the identification $N\cong \hat PNP$ for $N\in\cN_{\Sp(32,\bbR)}(\PSO(6))$.
We arrive at the space
\begin{align}\label{SN6ugly}
\fS^{\cN=6}_{\SO(6)}\ &=\ 
\left[\big(\,
\bbR^+\backslash\PSL(2,\bbR)_0 \times \SL(2,\bbR)\, \big)
 / \U(1)\right]
 / \bbZ_2^{\pP}   
\end{align}
where $\bbR^+$ corresponds to rescalings of the ungauged vector and the \U(1) is generated by \eqref{tU1}.
This space is parameterized by the gauge coupling constant, shifts in the theta-angles for the \SO(6) and the ungauged vectors, and a duality rotation linearly independent from \eqref{tU1}.
The latter can be taken to be the ${\omega}$-deformation acting uniformly on all sixteen vector fields, or equivalently an electric-magnetic rotation of the ungauged vector only.
The space \eqref{SN6ugly} can be parameterized by the double-coset representative
\begin{formula}
\label{SN6matrix}
S_{\cN=6}({\lambda},{\omega},{\theta}_{\SO(6)},{\theta}_0) = 
\begin{pmatrix}
{\lambda}\mathbf1_{15}   & 0                              &  \frac{{\theta}_{\SO(6)}}{2{\pi}}\mathbf1_{15} & 0                                 \\
             0   & \cos4{\omega}-\frac{{\theta}_0}{2{\pi}}\sin4{\omega}   &  0                                  & \sin4{\omega}+\frac{{\theta}_0}{2{\pi}}\cos4{\omega}      \\
   {\bf0}_{15}   & 0                              & \frac1{\lambda}{\bf1}_{15}                  & 0                                 \\
             0   & -\sin4{\omega}                        & 0                                   & \cos4{\omega}
\end{pmatrix}.
\end{formula}
Here ${\lambda}>0$ induces a rescaling of the gauge coupling constant, ${\theta}_{\SO(6)}$ is a shift in the theta-term for the gauged vectors, ${\theta}_0$ acts similarly for the singlet vector, and ${\omega}$ is the ${\omega}$-deformation inherithed from the maximal theory.
It acts only on the singlet vector because we exploited the quotient by the action of \eqref{tU1}.
The periodicity of ${\omega}$ is still the same as in $\cN=8$, namely ${\omega}\cong{\omega}+k\frac{\pi}4$, while parity induces the further identification
\begin{formula}
({\lambda},{\omega},{\theta}_{\SO(6)},{\theta}_0)\ {\cong}\ ({\lambda},-{\omega},-{\theta}_{\SO(6)},-{\theta}_0).
\end{formula}
Notice that, even setting ${\theta}_{\SO(6)}={\theta}_0=0$, parity as a \emph{local} symmetry is broken unless ${\omega}=0\ \mathrm{mod}\,\frac{\pi}8$.
For any other value of ${\omega}$ the action of parity must be combined with a \U(6) transformation and a symplectic dualization of the ungauged vector in order to preserve the Lagrangian.
For ${\omega}={\pi}/8$ the intrinsic parity of $A_{\mu}^0$ is reversed.

Compared to maximal supergravity, if we define our gauged models in the electric frame and only quotient by local field redefinitions, $\cN=6$ \SO(6) gauged supergravity actually admits one more deformation parameter associated with the difference in the theta-terms of the gauged and singlet vector potentials.\footnote{The difference in theta-terms can be accommodated in maximal supergravity by also adding extra Chern–Simons-like terms, following the general embedding tensor formalism.}
Notice that these shifts and ${\omega}$ are entirely on the same footing, as they do not affect the equations of motion of $\cN=6$ supergravity, and can be removed by a change of symplectic frame that leaves $`X0_MN^P $ invariant.
This also means that an uplift to IIA supergravity in $\CP^3$ is always possible, after the obvious dualizations similar to the one described in \cite{Borghese:2014gfa} for ${\omega}$.

\subsection{Deformations of $\cN=2$ $\SO(6)\times\SO(2)$ gauged supergravity}

The crucial difference when studying the $\cN=2$ gauged supergravity twin of the $\cN=6$ model is that, truncating form the $\cN=8$ \SO(8) theory, a FI term gauging $\SO(2)\subset\SU(2)_R$ is also induced \cite{Andrianopoli:2008ea}.
This means that the \SO(6) singlet vector $A_{\mu}^0$ is minimally coupled to the fermions of the $\cN=2$ theory, and the embedding tensor contains an extra vector ${\xi}_M = g_{\rm FI} {\delta}^0_M$ parameterizing this coupling.\footnote{We do not need to specify the explicit embedding of \SO(2) in $\SU(2)_R$ for our computation. Also notice that at the linearized level for vanishing scalar fields $A_{\mu}^0$ is the graviphoton.}
From this observation it is natural to anticipate that the ${\omega}$ parameter induced from $\cN=8$ will now be entirely non-trivial.
We also find an extra non-trivial deformation associated with  the ratio of the \SO(6) and FI gauge couplings which breaks compatibility with maximal (and $\cN=6$) gauged supergravity.

The computation of the space of deformations is analogous to the $\cN=6$ case. 
The only difference with respect to \eqref{SredQuotientN6} is in the left quotient, where now also ${\xi}_M$ must be stabilized \emph{up to a sign flip} that can be reabsorbed in an $\SU(2)_R$ transformation according to \eqref{bbS}.
Hence the new left quotient will be
\begin{formula}
\mathbb S(`X0_MN^P ,\, `{\xi} ) = \bbR_{{\theta}_{\rm \SO(6)}}\times\bbR_{{\theta}_0} \times \bbZ_2^2
\end{formula}
where one $\bbZ_2$ factor is associated with the center of $\SL(2,\bbR)_0$ and the other one is the outer automorphism of \PSO(6).
The two real lines are shifts in the theta parameters.
We immediately arrive at the new deformation space 
\begin{formula}\label{SredN2}
(\fS_{\rm red})^{\cN=2,}_{\SO(6)\times\SO(2)}\ &=\ 
\left[(\bbR_{{\theta}_{\rm \SO(6)}}\times\bbR_{{\theta}_0})\ \backslash\ (\PSL(2,\bbR)_0\times\SL(2,\bbR))\ /\ \U(1) \right]\,/\,\bbZ_2^\pP\ .
\end{formula}
The full $\fS$ space also keeps the theta-angle shifts.
The space \eqref{SredN2} is parameterized by the symplectic transformation
\begin{formula}\label{SN2matrix}
S_{\cN=2}({\lambda},{\lambda}_0,{\omega}) = G_{{\lambda}_0} S_{\cN=6}({\lambda},{\omega},0,0),
\end{formula}
where $G_{{\lambda}_0} = \mathrm{diag}({\bf1}_{15},\,{\lambda}_0,\,{\bf1}_{15},\,\frac1{{\lambda}_0})$, ${\lambda}_0>0$ rescales the FI coupling constant.
The periodicity of ${\omega}$ and the parity identifications are unchanged, hence in particular for the classically relevant deformations we have the fundamental domain
\begin{formula}
g_{\SO(6)}>0,  \qquad  g_{\rm FI}>0,  \qquad  {\omega}\in[0,\tfrac{\pi}8].
\end{formula}
where we traded the rescalings for the gauge coupling constants themselves.
In contrast to $\cN=6$, beyond an overall rescaling of all gauge couplings we have two non-trivial parameters, one being the ${\omega}$-deformation inherited from the truncation of maximal \SO(8) gauged supergravity, and the other genuinely new, associated with the rescaling of the ratio $g_{\rm FI}/g_{\SO(6)}$.

\section[Half-maximal gauged supergravities]{Half-maximal gauged supergravities\footnote{I thank Adolfo Guarino for many inspiring and informative discussions on the gaugings of half-maximal supergravity.}}\label{sec4}

Historically, a first important example of inequivalent gaugings of extended supergravities sharing the same gauge group are certainly the de Roo\,--Wagemans angles in half-maximal supergravity \cite{deRoo:1985jh}, where they are crucial for the existence of vacuum solutions of the scalar potential.
It is therefore natural to make contact with these duality phases in our setting, and complement them with an analysis of other possible deformation parameters that may arise in some models.

Let us first briefly summarize the continuous and discrete symmetries of ungauged half-maximal supergravity.
Pure $\cN=4$ supergravity has an \SL(2,\bbR) duality symmetry, reflected in the \SL(2,\bbR)/\SO(2) non-linear sigma model described by the complex scalar (axio-dilaton) in the gravity multiplet.
Moreover, there is an \SU(4) symmetry under which the fermions transform in the fundamental and anti-fundamental representations and the vectors in the $\bf6$.
When we couple the gravity sector to $\nv$ vector multiplets, the scalar fields in the latter combined with the axio-dilaton parameterize the coset space
\begin{equation}
\frac{\SL(2,\bbR)}{\SO(2)}\times\frac{\SOp(6,\nv)}{\SO(6)\times\SO(\nv)}~.
\end{equation}
Notice that we took only the identity component of \SO(6,\nv), since only continuous isometries are necessary to construct the scalar manifold.
For our purposes however it is crucial to take into account also any discrete symmetries.
We notice that all fields in the vector multiplets including fermions carry a vector index under \SO(\nv), and it is clear by inspection of the Lagrangian (see e.g. \cite{SchonWeidner}) that all couplings are actually invariant under the full (non-linearly realized) \O(\nv) group, hence extending \SOp(6,\nv) to \Op(6,\nv).
Extending the \SO(6) factor to an internal \O(6) is instead not possible, because fermions transform in chiral spinorial representations.
This means however that the action of space-time parity induces the appropriate \O(6)/\SO(6) automorphism of \SU(4) on both fermions and bosons.
Thus a full realization of an \O(6,\nv) symmetry is obtained combined with space-time parity, which also acts non-trivially on \SL(2,\bbR) by inducing a sign flip of the axion in the gravity multiplet.
We will come back to these transformations in the next section and provide an explicit embedding in the symplectic group.

\subsection{From $\cN=8$ to $\cN=4$: frames and parity}
\label{N8toN4}

Since we will mostly focus on gauge groups related to truncations of gauged maximal supergravity, it is convenient to put some facts about the half-maximal theory coupled to six vector multiplets in the perspective of a truncation from $\cN=8$.

The truncation is obtained from a $\bbZ_2$ projection on the spectrum and embedding tensor of the maximal theory \cite{Aldazabal:2011yz,Dibitetto:2011eu}.
We can take the $\bbZ_2$ to be embedded in \SU(8) as the matrix $\,U=\mathrm{diag}(\mathbf1_4,-\mathbf1_4)\,$ in the fundamental representation, thus breaking \SU(8) to the local $\SU(4)_R\times\SU(4)\times\U(1)$ of the half-maximal theory.
By construction all surviving bosonic and fermionic fields only transform under $\SU(4)/\bbZ_2\simeq\SO(6)$ as expected, while $\SU(4)_R$ is faithfully represented on the fermions.
The former \SO(6) is actually extended to an \O(6) symmetry as discussed in the previous section.
If we consider the maximal theory in the standard \SL(8,\bbR) symplectic frame, the truncation to $\cN=4$ yields a Lagrangian which is only invariant under an $\SO(3)^4$ subgroup of \Op(6,6).
Full \Op(6,6) invariance is obtained after dualization of six vectors: three in the gravity multiplet and three in the matter multiplets (at the linearized level), associated respectively to one \SO(3) factor in $\SU(4)_R$ and one in \O(6).
For concreteness, let us write down the kinetic terms for the vector fields in the resulting symplectic frame:
\begin{formula}
\label{N4kinvect}
\cL_{\rm v} = -\frac14\left(\Im{\tau} M_{{\Lambda}{\Sigma}} `F_{\mu}{\nu}^{\Lambda} `F^{\mu}{\nu}{\Sigma} +\frac12\Re{\tau}\, {\eta}_{{\Lambda}{\Sigma}} `{\epsilon}_{\mu}{\nu}{\rho}{\sigma} `F_{\mu}{\nu}^{\Lambda} `F_{\rho}{\sigma}^{\Sigma} \right)~,
\end{formula}
where ${\tau}$ is the axio-dilaton field of the gravity sector, ${\eta}_{{\Lambda}{\Sigma}}$ is the \O(6,6) invariant metric and $M_{{\Lambda}{\Sigma}}$ is the \O(6,6) generalized metric constructed from the scalars in the vector multiplets.
As already emphasized the duality symmetries of the theory are
\begin{formula}
\label{N4dualsym}
\SL(2,\bbR)\times\Op(6,6)~,
\end{formula}
where only $\PSL(2,\bbR)=\SL(2,\bbR)/\bbZ_2$ is realized on bosons.
These transformations are embedded in \Sp(24,\bbR) as matrices
\begin{formula}
\label{N4dualityembed}
\SL(2,\bbR):&\quad
\begin{pmatrix}
a\mathbf1_{12} & -b{\eta} \\
-c{\eta} & d\mathbf1_{12}
\end{pmatrix},&&ad-bc=1~,\\[1ex]
\Op(6,6):&\quad\begin{pmatrix}
K & \\
 & K^{-T} \\
\end{pmatrix}, &&K\in \Op(6,6)~.
\end{formula}
It is evident that the centers of the two factors in \eqref{N4dualsym} are represented in the same way on the bosons, which is consistent with the $\bbZ_2$ quotient of the \SL(2,\bbR) factor.%
\footnote{Our conventions can be mapped into those of Schon--Weidner \cite{SchonWeidner} by raising the \SO(6,n_{\rm v}) index of the dual field strength: $`G_{\mu}{\nu}{\Lambda} \to `{\eta}^{\Lambda}{\Sigma} `G_{\mu}{\nu}{\Sigma} $.}

It will be necessary to also consider the action of parity on the vectors and scalar fields of the theory.
Starting from the maximal theory and before any dualization, all electric vectors have the same parity assignment.
This is reflected in the action of parity on the bosons of the maximal theory as the outer automorphism of \Eseven, realized as the anti-symplectic transformation \cite{Ferrara:2013zga,DallAgata:2014ita}
\begin{formula}
P = {\sigma}_3\otimes\mathbf1_{28}.
\end{formula}
Crucially, this transformation acts on \ESU coset representatives as \cite{DallAgata:2014ita}
\begin{formula}
\label{ParityCosetN8}
`P_{\bbM}^{\bbN} \cV(\phi)`{}_\bbN^IJ = [\cV^*(\pP\phi')]`{}_\bbM\,IJ ~,
\end{formula}
where $\bbM,\,\bbN$ are fundamental \Eseven indices, $I,J$ are \SU(8) ones, and $\pP\phi$ is the action of space-time parity on the spin-0 fields.
The complex conjugation is consistent with the exchange of chiralities of the fermions.
When we truncate to $\cN=4$ and dualize six vectors, parity induces again an outer automorphism of \eqref{N4dualsym}, realized as the anti-symplectic transformation
\begin{formula}
\label{N4parity}
P=\text{diag}(+{\bf1}_3,\,-{\bf1}_3,\,+{\bf1}_3,\,-{\bf1}_3\, | \,
-{\bf1}_3,\,+{\bf1}_3,\,-{\bf1}_3,\,+{\bf1}_3)~,
\end{formula}
where we emphasized the separation in electric and magnetic components.
Applied to \eqref{N4dualityembed}, $P$ induces a sign flip of $b,c$ consistent with ${\tau}\to-{\tau}^*$.
At the same time it affects the \Op(6,6) factor, acting as an \O(6,6) discrete transformation that reverses the orientation of both eigenspaces of ${\eta}_{{\Lambda}{\Sigma}}$ (i.e. the six `timelike' and six `spacelike' directions).
Since the scalar fields of the vector multiplets form the $({\bf6},{\bf6})$ of $\O(6)\times\O(6)$, we obtain that half of them are pseudo-scalars as should be expected from the truncation from $\cN=8$.
This further induces a transformation similar to \eqref{ParityCosetN8} for the \Op(6,6) coset representative:
\begin{formula}
\label{ParityCosetN4}
`P_{\Lambda}^{\Sigma} \cV(\phi)`{}_{\Sigma}^i = [\cV^*(\pP\phi)]`{}_{\Lambda}\,i ~,\qquad
`P_{\Lambda}^{\Sigma} \cV(\phi)`{}_{\Sigma}^A = [\cV(\pP\phi)]`{}_{\Lambda}^A ~,
\end{formula}
where we wrote the representatives in a mixed form with ${\Lambda},\,{\Sigma}$ fundamental \Op(6,6) indices, $i$ a fundamental index of $\SU(4)$ and $A$ is a vector index of \O(6).
Again, these transformation properties are consistent with the action of parity on fermion fields.

\subsection{Pure $\cN=4$ \SO(4) gauged supergravities}

As a warmup for the more complicated case of the next section, let us consider the \SO(4) gaugings of pure half-maximal supergravity \cite{Freedman:1978ra,Das:1977uy}.
In this case only the six vectors in the gravity multiplet are present, and they sit in the adjoint of the gauge group $\SO(4)\subset\SU(4)$, so that its embedding in the symplectic group is really as $\SO(3)_+\times\SO(3)_-$:
\begin{formula}
\Sp(12,\bbR)\ni
\begin{pmatrix}
k_+&&&\\
&k_-&&\\
&&k_+&\\
&&&k_-
\end{pmatrix}~,
\qquad
k_\pm \in \SO(3)_\pm~.
\end{formula}
We will consider as a reference gauging the one where the gauge connection matrix is electric and the gauge couplings of the two simple factors are both equal to 1:
\begin{formula}
`{\vartheta}0_{\Lambda}^x = \begin{pmatrix}
 {\bf1}_3 & \\
 & {\bf1}_3 \\
 \end{pmatrix},\qquad
 `{\vartheta}0^{\Lambda}x = 0,\qquad
\end{formula}
where $x$ runs in the adjoint of \SO(4) and ${\Lambda}$ enumerates the vectors and their duals.
The reference embedding tensor is $`X0_M = `{\vartheta}0_M^x t_x $, $t_x\in\SO(3)_+\times\SO(3)_-$.
The relevant normalizers and stabilizers are easily identified:
\begin{formula}
\label{pureN4normstab}
\cN_{\Sp(12,\bbR)}(\PSO(4))&=\SL(2,\bbR)^2\rtimes\bbZ_2~,\\
\cN_{\SO(6)\times\PSL(2,\bbR)_{\tau}}(\PSO(4))&=\bbZ_4\times\PSL(2,\bbR)_{\tau}
\simeq\bbZ_2\times\SL(2,\bbR)_{\tau},\\
\bbS(X^0)&=\bbR_{\theta}^2\rtimes\bbZ_2~.
\end{formula}
The \SL(2,\bbR) factors in the first row correspond to separate dualizations of the vectors gauging each \SO(3) group, while conjugation of the gauge group by $\bbZ_2$ exchanges the two factors.
Shifts in $\SO(3)^2$-invariant theta-angles for the electric vectors stabilize the embedding tensor and correspond to the $\bbR^2$ term in the third row.
Finally, the $\bbZ_4$ term acts like the $\bbZ_2$ above and also flips the sign of the gauge connection of one \SO(3) factor.
Acting with it twice just flips the signs of all six vectors, which can be conveniently regarded as an $\SL(2,\bbR)_{\tau}$ transformation at the bosonic level.
We arrive at the reduced space of deformations%
\begin{formula}
\label{pureN4coset}
\mathfrak S_{\rm red}\ =\ \left( \bbR^2\ \backslash\ \SL(2,\bbR)^2\ /\ \SL(2,\bbR)_{\tau}\right)\, /\, (\bbZ_2\times\bbZ_2^{\pP})\ ,
\end{formula}
where the $\bbZ_2$ exchange quotient acts from both the left and the right.\footnote{Actually, $\bbZ_2$ and $\bbZ_2^{\pP}$ do not commute the way they are realized.
Their commutator is the center of $\SL(2,\bbR)_{\tau}$ so it is trivialized anyway.}
It is immediate to see that by gauge fixing $\SL(2,\bbR)_{\tau}$ we can take as coset representative (in the fundamental representation)
\begin{formula}
\label{pureN4cosetrep}
L(g,{\alpha}) \equiv L_+ \oplus L_-(g,{\alpha}),\quad
L_+ = {\bf1}_2,\quad 
L_-(g,{\alpha})=\begin{pmatrix}
\frac1{g} & \\
& g
\end{pmatrix}\,
\begin{pmatrix}
\cos{\alpha}&\sin{\alpha}\\
-\sin{\alpha}&\cos{\alpha}
\end{pmatrix}~.
\end{formula}
The parameter ${\alpha}$ corresponds to an electric-magnetic phase for $\SO(3)_-$, while the natural interpretation of $g$ is discussed below depending on the value of ${\alpha}$.
It is exhaustive to take $g\neq0$ and ${\alpha}\in(-\frac{\pi}2,\,\frac{\pi}2]$.
The expression above still admits a continuous residual symmetry corresponding to a shift in the axion (reflected as an $\SL(2,\bbR)_{{\tau}}$ transformation) combined with an $\bbR$ transformation in the left quotient, that shifts the constant theta term for the $\SO(3)_+$ gauge group so that $ L_+ = {\bf1} $ remains invariant:
\begin{formula}
\label{pureN4resgaugefix}
L(g,{\alpha}) \to L(\tilde g,\tilde {\alpha}) =
(T_{-{\theta}} \, L_+ T_{\theta}) \oplus( T_{\tilde {\theta}} L_-(g,{\alpha}) T_{\theta})~,
\quad
T_{\theta} \equiv 
\begin{pmatrix}
1&\frac{\theta}{2{\pi}}\\
 &1
\end{pmatrix}
\end{formula}
The further shift $T_{\tilde{\theta}}$ is necessary to preserve the gauge-fixing of $L_+$ and is associated with a constant theta-term for $\SO(3)_-$.
When ${\alpha}\neq0$ we can use this transformation to fix ${\alpha}={\pi}/2$, so that ${\alpha}$ becomes a discrete parameter.

Let us discuss the interpretation of ${\alpha}$ assuming momentarily that $g>0$.
Its two values correspond to the two well-known gaugings \cite{Freedman:1978ra,Das:1977pu} of \SO(4) in pure half-maximal supergravity: for ${\alpha}=0$ the gauging is electric in the \SU(4) covariant symplectic frame of \cite{Cremmer:1977tt} and corresponds to the Freedman--Schwarz gaugings \cite{Freedman:1978ra}, arising as an $S^3\times S^3$ truncation of IIA supergravity.
For ${\alpha}={\pi}/2$ the gauging is electric in the original \SO(4) covariant frame of \cite{Das:1977uy,Cremmer:1977tc} and comes from a truncation of eleven-dimensional supergravity on a seven-sphere \cite{Cvetic:1999au}.
The interpretation of $g$ is slightly different in the two cases.
When ${\alpha}=0$, $g$ represents the ratio of the gauge couplings of the two \SO(3) factors, while an overall rescaling of the two can be absorbed in a redefinition of the dilaton field, associated with the diagonal element of $\SL(2,\bbR)_{\tau}$.
For ${\alpha}={\pi}/2$ the situation is reversed: the redefinition of the dilaton field  can be used to set the two gauge couplings equal to each other, so that $g$ can be mapped into the overall gauge coupling of the entire \SO(4) gauge group.
These findings reproduce the discussion of \cite{Cvetic:1999au}, where the ${\alpha}=0$ case was also related to a singular limit of the ${\alpha}={\pi}/2$ one, corresponding to deforming $S^7$ into $S^3\times S^3\times S^1$.
In our language, this limit is associated with \eqref{pureN4resgaugefix} where ${\alpha}\neq0$ is sent to zero by an infinite shift of the axion.

Let us now complete our analysis taking into account the sign of $g$ and the two $\bbZ_2$ identifications in \eqref{pureN4coset}.
The internal $\bbZ_2$ is the combination of the ones in the second and third rows of \eqref{pureN4normstab}. It exchanges $L_+$ and $L_-$ and flips the sign of the latter.
Once we go back to the gauge-fixing $L_+=\bf1$, the resulting transformation is
\begin{formula}
L_-(g,{\alpha})\to -L_-(g,{\alpha})^{-1}~,
\end{formula}
which induces $g\to-1/g$ for ${\alpha}=0$, while it leaves $L_-(g,{\pi}/2)$ invariant.
Finally, the parity identification reverses the signs of ${\alpha}$ and $g$, which means that it also acts trivially for ${\alpha}={\pi}/2$.%
\footnote{one can see that $P={\sigma}_3\otimes{\sigma}_3\otimes{\bf1_3}$ and $\hat P={\sigma}_3\otimes{\bf1_6}$, where from now on we will use $\otimes$ to denote Kronecker products.}
Putting everything together, we arrive at the following families of gaugings:
\begin{formula}
\begin{array}{lll}
{\alpha}=0  &  g\in(0,1] & \hphantom{|}g\hphantom{|}\text{ $\sim$ ratio of gauge couplings,} \\
{\alpha}={\pi}/2  &  g\in(0,+\infty) & \hphantom{|}g\hphantom{|}\text{ $\sim$ overall gauge coupling,} \\
{\alpha}={\pi}/2  &  g\in(-\infty,0) & |g|\text{ $\sim$ overall gauge coupling.} \\
\end{array}
\end{formula}
The first two cases were discussed above.
The third case differs from the second in the relative signs of the gauge couplings of the two simple factors \cite{Gates:1982ct}.
It can arise from a truncation of \SO(4,4) gauged maximal supergravity, and can be lifted to eleven-dimensional supergravity on a seven dimensional hyperboloid \cite{Hohm:2014qga,Baron:2014bya}.

Finally, if we define these gaugings in their electric frames and do not want to allow for extra boundary terms to be added to the action we cannot quotient out $\bbR_{\theta}^2$.
In this case ${\alpha}$ is continuous.

\subsection{The $\cN=4$ $\SO(4)\times\SO(3)^2$ supergravities}
Let us now consider the maximal compact gaugings of $\cN=4$ supergravity coupled to six vector multiplets, embedded in the global internal symmetry group $ \SL(2,\bbR) \times {\rm O}^+(6,6) $.
The gauge group is $\SO(4)\times\SO(3)^2$, where the first factor is embedded in $\SU(4)_R$, so that fermions transform in the vector representation of $\SO(4)\subset\SU(4)_R$.
The $\SO(3)^2$ factors are embedded in the \O(6) symmetry of the vector multiplets, so that fields in the gravity multiplet are invariant.

We take again as reference gauging the case where all simple factors are gauged electrically in the frame of \eqref{N4kinvect} and all gauge couplings are equal to 1.
The appropriate normalizers and stabilizers are
\begin{formula}
\label{N4normstab}
\cN_{\Sp(24,\bbR)}(\SO(3)^4)&=\SL(2,\bbR)^4\rtimes \bbS_4~,\\
\cN_{\Op(6,6)\times\PSL(2)_{\tau}}(\SO(3)^4)&=\bbZ_4\times \bbD_4\times\PSL(2,\bbR)_{\tau}
,\\
\bbS(X^0)&=\bbR_{\theta}^4\rtimes \bbS_4~.
\end{formula}
Analogously to the previous case, conjugation by $\bbS_4$ acts as permutations of the four simple factors, $\bbZ_4$ acts like in the previous section for $\SO(4)\subset\SU(4)_R$ and the dihedral group $\bbD_4$ acts similarly for $\SO(3)^2\subset\O(6)$ but also includes the sign flip of one gauge connection without any exchange.
Fixing the $\bbS_4$ quotient, we can then write the reduced $\mathfrak S$-space as
\begin{formula}
\mathfrak S_{\rm red}\ =\ \left( \bbR^4\ \backslash\ \SL(2,\bbR)^4\ /\ \PSL(2,\bbR)_{\tau} \right)\,/\,(\bbZ_4\times\bbD_4)\rtimes\bbZ_2^\pP\ .
\end{formula}
The discrete identifications act from both the left and the right side: $\bbZ_4\times\bbD_4$ is embedded in $\SO(6)\times\O(6)$ as natural, and its right action is combined with left $\bbS_4$ compensating transformations.
Parity acts as described in section~\ref{sec:parity}.

We can perform a first round of gauge fixing by defining the coset representatives as
\begin{formula}
\label{N4SO4square coset repr}
L(g,{\alpha},h_i,{\beta}_i) \equiv L_+ \oplus L_-(g,{\alpha}) \oplus L_1(h_1,{\beta}_1) \oplus L_2(h_2,{\beta}_2),
\end{formula}
where $L_+={\bf1}_2$ and $L_i(h_i, {\beta}_i)$ are defined as $L_-(g,{\alpha})$ in \eqref{pureN4cosetrep}.
Notice that we have also used the $-1_{12}$ element of $\bbZ_4\times\bbD_4$ to fix the sign of $L_+$.
Exploiting $\bbD_4$ we can already reduce the range of the parameters to
\begin{formula}
\label{N4initialrange}
g\neq0,\quad h_i>0,\quad {\alpha},\,{\beta}_i\in\left(-\tfrac{\pi}2,\tfrac{\pi}2\right],\quad
{\beta}_1>{\beta}_2\ \ \text{or} \ \ {\beta}_1={\beta}_2\ \&\ h_1\ge h_2~.
\end{formula}
The same combination of an axion and theta-angle shift we discussed for pure supergravity can be used to set to ${\pi}/2$ one non-vanishing angle.
Hence, we have three separate cases:
\begin{formula}\label{N4cases}
\begin{tabular}{ll}
Case 0:\  &${\alpha}={\beta}_i=0$,\\[.5ex]
Case I:\  &${\alpha}=0,\ {\beta}_1=\frac{\pi}2$, ${\beta}_2\in\left(-\tfrac{\pi}2,\tfrac{\pi}2\right]$,\\[.5ex]
Case II:\ &${\alpha}=\frac{\pi}2$, ${\beta}_i\in\left(-\tfrac{\pi}2,\tfrac{\pi}2\right]$.
\end{tabular}
\end{formula}

We now must consider the action of the $\bbZ_4$ discrete quotient acting on $L_\pm$.
Enforcing the gauge-fixing of $L_+$ and the restriction on the ranges of the angles ${\beta}_i$, its action is reduced to just a $\bbZ_2$.
We obtain the following transformation rules:
\begin{formula}
&\text{Case 0:}\ && g\to-\frac1g,\ \ h_i\to\frac{h_i}{|g|}~,\\[1ex]
&\text{Case I:}\ && g\to-\frac1g,\ \ h_1\to |g| h_1,\\&&&
h_2 \to |g|\, h_2\, (g^4\cos^2{\beta}_2+\sin^2{\beta}_2)^{-1/2}\,, 
\ \ {\beta}_2\to \arctan\left(\frac1{g^2}\tan{\beta}_2\right)
~,
\\[1ex]
&\text{Case II:}\ && g\to g,\ \ 
h_i \to |g|\, h_i\, (g^4\cos^2{\beta}_i+\sin^2{\beta}_i)^{-1/2},\ \ 
{\beta}_i \to -\arctan(g^2\cot{\beta}_i)~.
\end{formula}
We regard the transformation rules of the angles to be normalized, e.g. ${\beta}_2={\pi}/2\to{\pi}/2$ for Case~I and ${\beta}_i=0\to{\pi}/2$ for Case~II.
Of course $\bbD_4$ allows us to exchange $i=1$ and $i=2$ if necessary to guarantee that \eqref{N4initialrange} is satisfied.

The only remaining identification is parity. 
For convenience, we will combine its action with the reflection element of $\bbD_4$, so that all (electric) vectors in the vector multiplets have positive eigenvalue under $P_M{}^N$.
Therefore, embedding our coset representative in \Sp(24,\bbR) we must quotient by
\begin{formula}\label{blah1}
L(g,{\alpha},h_i,{\beta}_i) \simeq \hat P\, L(g,{\alpha},h_i,{\beta}_i)\, P~,
\end{formula}
where $P={\sigma}_3\otimes\text{diag}(+{\bf1}_3,\,-{\bf1}_3,\,+{\bf1}_3,\,+{\bf1}_3)$ and $\hat P={\sigma}_3\otimes{\bf1}_{12}$\,.
We arrive at the identifications
\begin{formula}
&\text{Case 0:}\ && g\to-g~,\\[.5ex]
&\text{Case I:}\ && g\to-g~,\ \ {\beta}_2\to-{\beta}_2~,\\[.5ex]
&\text{Case II:}\ && {\beta}_i\to-{\beta}_i ~.
\end{formula}

We can now provide the full classification of inequivalent $\SO(4)\times\SO(3)^2$ gaugings of half-maximal supergravity coupled to six vector multiplets.
One simple factor in the gravity sector is always gauged by vectors that are electric in the \SU(4) frame, and with gauge coupling equal to one.
To each other factor is associated a gauge coupling and a duality phase described above, whose values are further restricted by \eqref{N4cases}.
The complete fundamental domain is:
\begin{formula}\label{N4fundom}
&\text{Case 0:}\ && g\in(0,1]~,\ h_1\ge h_2>0 ~,\\[1ex]
&\text{Case I:}\ && g\in(0,1]~,\ h_i>0 \text{ and } {\beta}_2\in(-\tfrac{\pi}2,\tfrac{\pi}2)~,\text{ or }\ h_1\ge h_2>0 \text{ and } {\beta}_2=\tfrac{\pi}2~,\\[1ex]
&\text{Case II:}\ && g\neq0,\ \ a\equiv\arctan|g|\ \text{and}\\
&&&
-a\le{\beta}_2<{\beta}_1\le\tfrac{\pi}2~,\ \text{and} \ h_i>0~,\text{ or }
-a\le{\beta}_2={\beta}_1\le0 \text{ and } h_1\ge h_2>0~.
\end{formula}

This space of deformations is clearly more complicated than any of the examples discussed so far.
It can help to identify some familiar deformations among the (up to) five we have discovered.
The most natural question is what values of the parameters correspond to the truncation of maximal \SO(8) gauged supergravity, and whether its ${\omega}$ deformation survives.
Following the truncation and dualization procedure we discussed in section~\ref{N8toN4}, we conclude that ${\omega}$ survives and the truncations of the $\SO(8)_{\omega}$ gaugings are implemented as
\begin{formula}
g=h_i=1,\ \  {\beta}_1=\tfrac{\pi}2 -2{\omega},\ \ {\beta}_2=-2{\omega},\ \ {\omega}\in[0,\tfrac{\pi}8],
\end{formula}
which fits into Case II.
Moreover, if we flip the sign of $g$ we obtain truncations of the ${\omega}$-deformed \SO(4,4) maximal supergravities \cite{DallAgata:2012sx,DallAgata:2014ita}.

\subsection{$\SO(4,\bbC)$ gaugings}

As another example we can consider the gauging of 
\begin{formula}
\SOp(3,1)^2 \subset\SOp(3,3)^2\subset\SOp(6,6)~.
\end{formula}
also considered in \cite{deRoo:2003rm,Roest:2009tt}.
Taking into account that fermions transform in the double cover, the gauge group is really $\SO(4,\bbC)$, but we will still refer to it as $\SOp(3,1)^2$ to be able to distinguish the two factors.
This gauge group also arises as a truncation of the $\SO(4,4)_{\omega}$ maximal gauged supergravities.
First, the usual normalizers and stabilizers are
\begin{formula}
\cN_{\Sp(24,\bbR)}(\SOp(3,1)^2)&=\SL(2,\bbC)^2\rtimes \bbD_4~,\\
\cN_{\Op(6,6)\times\PSL(2)_{\tau}}(\SOp(3,1)^2)&=\bbZ_2\times \bbD_4\times\PSL(2,\bbR)_{\tau}
\simeq \bbD_4\times\SL(2,\bbR)_{\tau}~,\\
\bbS(X^0)&=\bbR_{\theta}^4\rtimes \bbD_4~.
\end{formula}
This time, conjugation of the gauge group by $\bbD_4$ forms the outer automorphism group of $\SOp(3,1)^2$, generated by exchange of the two factors and the outer automorphism on one.
For definiteness, if we embed the first (second) \SOp(3,3) in \Op(6,6) as acting on the first (last) three positive and negative eigenvectors of ${\eta}_{{\Lambda}{\Sigma}}$, then
$\bbD_4$ can be conveniently expressed as generated by the matrices
\begin{formula}
(\ii {\sigma}_2\oplus{\sigma}_1)\otimes{\bf1}_3~,\qquad({\bf1}_2\oplus{\sigma}_3)\otimes{\bf1}_3~,
\end{formula}
further embedded in the symplectic group.
Fixing the left $\bbD_4$ we can write the quotient as 
\begin{formula}
\label{SL2Cdefo}
\mathfrak S_{\rm red}\ =\ 
\left( \bbC^2\ \backslash\ \SL(2,\bbC)^2\ /\ \SL(2,\bbR)_{\tau} \right)\,/\,(\bbD_4\rtimes\bbZ_2^\pP)~,
\end{formula}
where the remaining dihedral group acts from both the left and the right side of the $\SL(2,\bbC)^2$ deformation matrices.\footnote{The left action is the transpose of the right action.}

Let us discuss the parameterization of the deformations for a single $\SO(3,1)$ factor of the gauge group.
The {deformation} \SL(2,\bbC) associated with it generally has six parameters, of which two correspond to shifts in theta-angles that we are ignoring.
It helps to regard the other four as deformations of the gauge connection rather than of the symplectic frame, in order to get a more intuitive interpretation of their effect.
Our initial consistent choice of gauge connection is that in which the \SO(3) subgroup is gauged by three of the vectors $A_{\mu}^{i-}$ associated with negative eigenvalues of ${\eta}_{MN}$ and the boosts are gauged by vectors $A_{\mu}^{i+}$ with positive eigenvalue, with $i=1,2,3$.
This means that in a pure supergravity truncation, where only $A_{\mu}^{i-}$ survive, the \SO(3) group would be gauged.
We also set the initial coupling constant to unit value.
Introducing also dual vector fields $A_{\mu}{}_{i\pm}$, the action of a generic \SL(2,\bbC) deformation yields the consistent gauge connections
\begin{formula}
\label{SL2Cconn1}
\SO(3):&\qquad 
 g\cos{\alpha}\,(\cos{\varphi} A_{\mu}^{i-} + \sin{\varphi} A_{\mu}^{i+}) 
+g\sin{\alpha}\,(\cos{\psi} A_{\mu}{}_{i-} + \sin{\psi} A_{\mu}{}_{i+})\\
\text{boosts}:&\qquad
 g\cos{\alpha}\,(-\sin{\varphi} A_{\mu}^{i-} + \cos{\varphi} A_{\mu}^{i+}) 
+g\sin{\alpha}\,( \sin{\psi} A_{\mu}{}_{i-} - \cos{\psi} A_{\mu}{}_{i+})\ .
\end{formula}
We can of course interpret $g$ and ${\alpha}$ as the gauge coupling constant and a de Roo--Wagemans angle for the \emph{whole} \SO(3,1).
When ${\alpha}=0$, ${\varphi}$ plays the role of the deformation phase introduced in \cite{Roest:2009tt} in terms of sums and differences of gauge couplings for the \SO(3) and boost generators of \SO(3,1).
Notice however that when ${\alpha}\neq0$, ${\varphi}$ and ${\psi}$ are two independent parameters, and together the three angles parameterize an $S^3$, corresponding to the compact subgroup of \SL(2,\bbC).%
\footnote{An alternative parameterization is in terms of a single `Roest--Rosseel' angle and two de Roo--Wagemans phases: one for \SO(3) and another for the boost generators, despite the fact that they do not decompose into two commuting simple algebras.}
In terms of a $2\times2$ complex matrix representing \SL(2,\bbC), the deformations above can be taken to define the transformation (for $\cos{\alpha}\neq0$)%
\footnote{Recall that if $S\in \mathfrak S_{\rm red}$, it is $S^{-1}$ that acts on $X_{MN}{}^P$.
When $\cos{\alpha}=0$, a different fixing of the extra theta-term shifts is necessary, where the upper-right block is non-vanishing.}
\begin{formula}
\label{SL2Crepr1}
L_1 =
\begin{pmatrix}
\frac{e^{\ii{\varphi}}}{g\cos{\alpha}}  &  0   \\
-{e^{\ii{\psi}}}{g\sin{\alpha}}  & e^{-\ii{\varphi}} g\cos{\alpha}
\end{pmatrix} \in \SL(2,\bbC)\ .
\end{formula}

For this single \SO(3,1) factor in the gauge group we can carry out at least the fixing of the continuous elements of the quotients in \eqref{SL2Cdefo}.
This means that we want to compute a representative of the double quotient
\begin{formula}
\label{SL2Cdefo single factor}
\bbC\ \backslash\ \SL(2,\bbC)\ /\ \SL(2,\bbR)~.
\end{formula}
The left quotient corresponds to matrices of the form 
$\left(
\begin{smallmatrix}
1&z\\0&1
\end{smallmatrix}
\right)$
with $z\in\bbC$, which we have used in \eqref{SL2Crepr1} to set one entry to vanish.
Exploiting the quotients we arrive at three separate one-parameter branches for the representative $L_1$ of \eqref{SL2Cdefo single factor}.
In terms of \eqref{SL2Crepr1}, they correspond to:
\begin{formula}
\label{SL2Cbranches}
\text{1)\ }& {\varphi}\in[0,{\pi})\,,\ g=1\,,\ {\alpha}=0\,,  \vphantom{\frac{\pi}2}\\
\text{2)\ }& {\varphi}=0\,,\ \ {\psi}=\frac{\pi}2\,,\ g\cos{\alpha}=1\,,\ g\sin{\alpha}\equiv r\neq0\,,\\
\text{3)\ }& {\varphi}=\frac{\pi}2\,,\ \ {\psi}=0\,,\ g\cos{\alpha}=1\,,\ g\sin{\alpha}\equiv r\neq0\,.
\end{formula}

The second \SL(2,\bbC) group of deformations associated with the other \SOp(3,1) gauge group is treated similarly, but only a left $\bbC$ quotient must be performed, giving a coset representative $L_2 \in \bbC\backslash\SL(2,\bbC)$.
This means that four non-trivial deformation parameters survive and can be parameterized just analogously to \eqref{SL2Cconn1}, with a new gauge coupling constant and new deformation phases.
The vectors in the gauge connection are of course the remaining six (which we can regard as $A_{\mu}^{i\pm}$ for $i=4,5,6$, together with their duals).

Once all continuous deformations and identifications have been taken into account, we should consider the discrete ones.
The generator of $\bbZ_4\subset\bbD_4$ acts on $L_1\otimes L_2$ as
$
L_1\otimes L_2  \to L_2^*\otimes L_1
$, while the reflection element can be taken to act as $L_2\to L_2^*$.
Finally parity, embedded as described below \eqref{blah1}, flips the signs of the off-diagonal elements of $L_1,\,L_2$ and conjugates $L_2$.

The remaining step is to combine these discrete transformations with the fixing of continuous identifications performed above.
We refrain from doing so here, as the task is rather complicated and the resulting transformations written in terms of the parameters $g,\,{\alpha},\,{\varphi},\,{\psi} $ of each gauge group turn out to be quite uninformative.
However, we stress that our analysis is sufficient to conclude that the $\SL(2,\bbC)^2$ gauging of half-maximal supergravity admits five non-trivial deformation parameters, separated in at least three branches as described in \eqref{SL2Cbranches}.

\subsection{Two examples outside $\SL(4,\bbR)^2$}

The gauge groups of the last two sections are embedded in a subgroup $\SOp(3,3)^2\sim\SL(4,\bbR)^2$ of \Op(6,6).
There are many more gauge group available in this sector, as studied e.g. in \cite{deRoo:2003rm,Roest:2009tt}.
There are also several different real forms of $\SO(4)^2$ that can be constructed.
If we investigate the embedding of such gauge groups in \Op(6,6) directly, we find eight possibilities (we ignore the details about the centers for brevity):
\begin{formula}
\label{N4 SO4xSO4 real forms}
&\SO(4)^2,\qquad\ \SOp(3,1)^2,\qquad\ \SO(2,2)^2,\qquad\,\SO(4)\times\SO(2,2),\\
&\SO(4)\times\SOp(3,1),\quad\SOp(3,1)\times\SO(2,2),\quad\SOs(4)\times\SO(2,2)\ {\small(\times2)},
\end{formula}
where the last entry counts twice because it admits two inequivalent embeddings in \Op(6,6).
Except for this last entry, the other groups are embedded in $\SOp(3,3)^2$ together with many group contractions.
A full classification of these groups and their deformations goes beyond the scope of this paper, but the two gaugings of $\SOs(4)\times\SO(2,2)$ are a curious example as they do not sit in $\SOp(3,3)^2$.
We have checked explicitly that they satisfy the embedding tensor constraints.

One embedding of this gauge group really gives rise to $\Gg=\SU(2)\times\SOp(1,2)^3$, where the compact factor is contained in the \SO(6) subgroup of \Op(6,6), acting on fermions as $\SU(2)\subset\SU(4)_R$.
The compact subgroups of $\SOp(1,2)^3$ are embedded in \O(6).
The procedure to construct the $\mathfrak S_{\rm red}$ space is analogous to the compact case, except that now the outer automorphism group of $\Gg$, reflected in $\cN_{\Sp(24,\bbR)}(\GX)$, is $\bbS_3\ltimes\bbZ^3_2$ accounting for permutations of $\SOp(1,2)^3$ and for the outer automorphism of each factor.
After some simplifications we obtain
\begin{formula}
\mathfrak S_{\rm red}=
\frac{\PSL(2,\bbR)^3}{\bbR^4\times\bbS_3},
\end{formula}
where the numerators are dualizations of each \SOp(1,2) factor and one $\bbR$ is the combination of axion and theta-term shift that can be used to fix a duality phase to $0$ or ${\pi}/2$.
This space can be parameterized in terms of a gauge coupling and a duality phase for each simple factor, following \eqref{N4SO4square coset repr}.
We take $L_+$ there to correspond to the \SU(2) factor here, so that it is gauged electrically with coupling equal to one. 
The parameters $g,\,{\alpha},\,h_i,\,{\beta}_i$ now are associated with each of the three non-compact factors.
Their fundamental domain is now given by \eqref{N4initialrange} and \eqref{N4cases} with the further condition $g>0$.

The other embedding is obtained when the compact factor is contained in \O(6).
This results in $\Gg=\SOp(2,2)\times\SOp(1,2)\times\SO(3)$.
The parameterization and fundamental domain for its duality orbits are analogous to the previous case, with the further restrictions
\begin{formula}
\begin{tabular}{ll}
Case I:\  &${\beta}_2\in\left[0,\tfrac{\pi}2\right]$,\\[.5ex]
Case II:\ & if ${\beta}_1={\beta}_2={\beta}$, then ${\beta}\in\left[0,\tfrac{\pi}2\right]$.
\end{tabular}
\end{formula}

\section{Gaugings of the STU model}\label{sec5}

As a final application we classify all gaugings of $\cN=2$ supergravity coupled to three vector multiplets and prepotential 
\begin{formula}
\label{mSTUprepot}
F(X) = -2i \sqrt{X^0 X^1 X^2 X^3}\ .
\end{formula}
This prepotential defines the so-called `magnetic' STU model, related to the standard cubic prepotential $F(X)\sim\frac{X^1 X^2 X^3}{X^0}$ by a change of symplectic frame.
The magnetic STU prepotential is more natural when discussing the theory as it arises from  a $\U(1)^4$-invariant consistent truncation of maximal supergravity.
This model is often the starting point for the construction of black hole solutions with both Minkowski and AdS asymptotics.

The three complex scalar fields parameterize three copies of the coset space $\SL(2,\bbR)/\U(1)$.%
\footnote{In terms of the square root prepotential, a good parameterization of the scalars is $z^i \equiv F_i(X)/X^0$, $i=1,2,3$.
In this case each scalar parameterizes one $\SL(2,\bbR)/\U(1)$ manifold.}
The internal global symmetries of the theory are the direct product of the duality symmetry $\SL(2,\bbR)^3\rtimes\bbS_3$ and the \SU(2) external automorphism of the supersymmetry algebra.
The permutation group $\bbS_3$ acts as triality \cite{hep-th/9508094} on the three vector multiplets and the associated \SL(2,\bbR) factors.  
Finally, the four vector fields and their duals transform in the $(\mathbf2,\mathbf2,\mathbf2,\mathbf1)$ of $\SL(2,\bbR)^3\times\SU(2)$.
We can denote them as $`A_{\mu}^a{\alpha}{\dot{\alpha}} $, where $a,{\alpha},\dot{\alpha}$ are fundamental indices for each of the factors $\SL(2,\bbR)_i$, $i=1,2,3$.
The electric vector fields in the symplectic frame of \eqref{mSTUprepot} are
\begin{formula}
(A_{\mu}^{0},\ A_{\mu}^{1},\ A_{\mu}^{2},\ A_{\mu}^{3})\equiv
(A_{\mu}^{111},\ -A_{\mu}^{122},\ -A_{\mu}^{212},\ -A_{\mu}^{221})\,,
\end{formula}
so that only a $\GL(1,\bbR)^3\rtimes\bbS_3$ subgroup of the duality symmetries is realized locally.\footnote{These identifications can be obtained e.g. from the transformation properties of the gauge-kinetic function, written in terms of the second derivatives of the prepotential $F_{{\Lambda}{\Sigma}}(X) $, ${\Lambda}=0,1,2,3$.}

The allowed gauge groups of the STU model are easily identified by the requirement that the vector fields transform in the adjoint representation and by imposing the linear constraint on the candidate embedding tensor.
They are FI-gaugings of $\U(1)\subset\SU(2)$, the gauging of a diagonal combination of two \SL(2,\bbR) groups (which form the electric group of the Lagrangian in an appropriate frame) and the combination of the two previous options.

\subsection{Fayet--Iliopoulos gaugings}

The simplest gauging of the STU model involves one linear combination of the vectors $A_{\mu}^{a{\alpha}\dot{\alpha}}$ gauging a $\U(1)\subset\SU(2)$.
The only charged fields under the gauge group are thus the fermions, so that the bosonic Lagrangian is only affected by the introduction of a scalar potential.
The whole gauging is determined by a constant moment map ${\xi}_{a{\alpha}\dot{\alpha}}$.
Technically, this is a triplet of \SU(2), but since only a single \U(1) can be gauged, and there is a unique embedding in \SU(2), we shall ignore the adjoint index in ${\xi}_{a{\alpha}\dot{\alpha}}$.
In our current symplectic frame explicit $\SL(2,\bbR)^3$ covariance is broken, and the eight entries of the FI term take the values
\begin{formula}
{\xi}_M = ({\xi}_{111}, -{\xi}_{122}, -{\xi}_{212}, -{\xi}_{221} \,|\  
       {\xi}_{222}, -{\xi}_{211}, -{\xi}_{121}, -{\xi}_{112} )~,
\end{formula}
where $M$ is as usual a symplectic index, mirroring the eight vectors $A_{\mu}^M$.

Classifying all possible FI gaugings is equivalent to classifying the symplectic deformations of a reference one, since the gauge group is always fixed.
The usual double quotient then would read:
\begin{formula}
\label{mSTU Sred}
\fS_{\rm red}=\left({\rm ISp}(6,\bbR)\ \backslash\ \Sp(8,\bbR)\ /\ \SL(2,\bbR)^3\right)\,/\,(\bbS_3\times\bbZ_2^\pP)~.
\end{formula}
In fact, constructing this double coset is entirely equivalent to explicitly building the $\SL(2,\bbR)^3\rtimes (\bbS_3\times \bbZ_2^\pP)$ orbits of the FI term itself, especially so because coset representatives of ${\rm ISp}(6,\bbR)\backslash\Sp(8,\bbR)$ are also entirely specified by a symplectic vector in the $({\bf 2},{\bf 2},{\bf 2})$.

The classification of duality orbits for a general FI term ${\xi}_M$ can be computed rather straightforwardly.
One approach is to break covariance with respect to one of the \SL(2,\bbR) factors and write two matrices ${\xi}_{1{\alpha}\dot{\alpha}}$ and ${\xi}_{2{\alpha}\dot{\alpha}}$.
The action of $\bbS_3$ is generated by transposition of these matrices, and exchange of the last row of ${\xi}_{1{\alpha}\dot{\alpha}}$ with the first of ${\xi}_{2{\alpha}\dot{\alpha}}$.
Then one can rely on the computation of the determinant of ${\xi}_1$ to set it to a reference form and use the non-covariant \SL(2,\bbR) to make ${\xi}_2$ orthogonal to ${\xi}_1$.
The remaining necessary steps are just making use of the conjugacy classes of $\fsl(2,\bbR)$ and taking into account further identifications given by triality and parity.
We find the following inequivalent FI gaugings:
\def\hpg{\hphantom{g}}
\begin{formula}
{\xi}_M = g&(1,1,1,1\,|\ 0,0,0,0) && \text{Anti de Sitter} &&\SO(8)_{\omega} \\
{\xi}_M = g&(-1,-1,1,1\,|\ 0,0,0,0) && \text{de Sitter} &&\SO(4,4)_{\omega} \\
{\xi}_M = \hpg&(1,0,0,0\,|\ 0,0,0,0) && \text{Minkowski}  &&\text{Scherk--Schwarz, }\SO(6,2)_{{\pi}/4},\, \ldots \\
{\xi}_M = g&(-1,1,1,1\,|\ 0,0,0,0) &&  &&\SO(6,2)_{\omega},\ {\omega}\neq{\pi}/4 \\
{\xi}_M = \hpg&(1,1,1,0\,|\ 0,0,0,0) &&  &&\CSO(6,0,2),\ \CSO(4,2,2) \\
{\xi}_M = \hpg&(1,1,0,0\,|\ 0,0,0,0) &&  &&\CSO(4,0,4) \\
{\xi}_M = \hpg&(-1,1,0,0\,|\ 0,0,0,0) &&  &&\CSO(2,2,4) \\
\end{formula}
We have indicated what kind of vacuum (if any) can be found in the six-dimensional scalar field space of the resulting theories, and \emph{some} of the uplifts of these models to gauged maximal supergravities.
Notice that it is always possible to find a representative of each orbit that is fully electric.
We can also see that the only continuous parameter allowed is a (positive) gauge coupling $g$, and even that is only available for certain orbits.

A particular choice of FI term gives rise to a gauged supergravity with a fully supersymmetric AdS vacuum, arising as a $\U(1)^4$-invariant consistent truncation of \SO(8) gauged maximal supergravity.
It is interesting to ask whether the ${\omega}$-deformation of the latter remains non-trivial in the truncation.
It was pointed out in \cite{Catino:2013ppa} that the scalar potential of the FI-gauged STU model resulting from the $\U(1)^4$ truncation of the \SO(8) maximal supergravities is independent from ${\omega}$ (also including the axions).
However, \cite{Lu:2014fpa} argued that the ${\omega}$-deformation is preserved in the truncation to STU, because the vector fields couple minimally to the fermions, and it is not possible to cancel ${\omega}$ by a duality.
Here we show that the ${\omega}$ deformation is in fact trivial in the STU model (up to boundary terms), since there are no continuous duality orbits for the FI term, except for the choice of gauge coupling constant.
The disagreement with \cite{Lu:2014fpa} is due to the fact that, even if a certain linear combination of vector fields is minimally coupled to fermions and thus cannot be freely dualized, three other vector fields are ungauged.
Symplectic redefinitions of the latter are therefore available and turn out to be sufficient to reabsorb ${\omega}$ from all couplings.

To show this explicitly, we take the point of view in which the deformation is entirely contained in the choice of embedding tensor, so that any $\mathrm{ISp}(6,\bbR)$ redefinition of the ungauged vectors is automatically taken into account.
After truncation of maximal \SO(8) gauged supergravity to the STU field content, the prepotential is still \eqref{mSTUprepot} and the electric and magnetic FI term takes the form
\begin{formula}
\label{FI omega}
{\xi}_M \propto (\cos{\omega},\,\cos{\omega},\,\cos{\omega},\,\cos{\omega}\,|\,\sin{\omega},\,\sin{\omega},\,\sin{\omega},\,\sin{\omega} )~.
\end{formula}
This is the most general FI term that gives rise to a fully supersymmetric AdS vacuum at the origin of scalar-field space.
We can remove ${\omega}$ e.g. by the $\SL(2,\bbR)_2$ rotation
\begin{formula}
\begin{pmatrix}
\cos{\omega} & & &\sin{\omega} \\
& \cos{\omega} &\sin{\omega} & \\
&-\sin{\omega} & \cos{\omega} & \\
-\sin{\omega}& & & \cos{\omega}
\end{pmatrix}\otimes{\bf1}_2\ .
\end{formula}
We thus conclude that:
\begin{itemize}
\item when defined in terms of an electric and magnetic FI term in a fixed frame, the ${\omega}$-deformation is entirely reabsorbed in an $\SL(2,\bbR)^3$ duality transformation;
\item when defined in an electric frame as done in \cite{Lu:2014fpa}, the ${\omega}$ deformation can be reabsorbed in a redefinition of the scalar fields that mirrors the \SL(2,\bbR) transformation above, combined with an $\mathrm{ISp}(6,\bbR)$ symplectic redefinition of the three ungauged vectors, compatibly with \eqref{mSTU Sred}.
\end{itemize}
This result holds for both the FI gaugings arising from truncations of the $\SO(8)_{\omega}$ and $\SO(4,4)_{\omega}$ maximal gauged supergravities.
Alternatively, for the AdS case the duality rotation can be taken in the \U(1) of the diagonal subgroup of $\SL(2,\bbR)^3$, showing that ${\omega}$ is trivial also in the `$3+1$' truncation of the STU model.

Since ${\omega}$ is trivial, we should ask how the BPS conditions of the black hole described in \cite{Lu:2014fpa} turned out to be ${\omega}$-dependent.
To answer it is sufficient to notice that if we stay in the symplectic frame defined by \eqref{mSTU Sred} with ${\omega}$-dependent FI term \eqref{FI omega}, both the bosonic Lagrangian and the solution of \cite{Lu:2014fpa} are completely ${\omega}$-independent.
The supersymmetry variations of the fermions, however, are affected by ${\omega}$, which explains why the supersymmetry of the black hole depends on the parameter.
Acting with an $\SL(2,\bbR)^3$ duality as just discussed, we can remove ${\omega}$ from the FI term, but now both the vectors and the six scalar fields of the non-supersymmetric black hole solutions will take a different form compared to the supersymmetric one.
They are therefore inequivalent field configurations of the same theory.

If we consider the STU truncation of $\SOs(8)_{\omega}\simeq\SO(6,2)_{\omega}$, the ${\omega}={\pi}/4$ model belongs to a different orbit than all other cases with ${\omega}\in [0,{\pi}/4)$.
Indeed, a Minkowski vacuum is found for ${\omega}={\pi}/4$ in the $\SO(6,2)$ maximal theory, with a moduli space that matches the STU model scalar sector \cite{DallAgata:2011aa,Catino:2013ppa}.\footnote{More precisely, there are several `STU' branches in the moduli space associated with how one breaks the gauge group to a Cartan subgroup.}
A huge family of gaugings with Minkowski vacua was found in \cite{Catino:2013ppa} starting from this gauging and taking singular limits in its moduli space, with $\cN=0,2,4,6$ residual supersymmetry.
They include as particular cases the Scherk--Schwarz and Cremmer--Scherk--Schwarz gaugings.
All these theories fall in the same duality orbit when truncated to a FI gauging of the STU model.

Finally, let us comment on the full $\fS$-space for the FI gaugings.
As should be clear from the previous discussion, this space corresponds to first choosing one of the seven conjugacy classes described above, and then deforming the Lagrangian by a symplectic redefinition of the three ungauged vectors (plus theta-shifts in the gauged one) that sits in the coset space
\begin{formula}
\frac{\rm ISp(6,\bbR)}{\GL(3,\bbR)\ltimes \bbR^3}
\end{formula}
where both the numerator and denominator stabilize ${\xi}_M$.
Furthermore, any duality, triality and parity symmetries that stabilize ${\xi}_M$ must be quotiented out.

\subsection{Charge quantization}

The Abelian gaugings of the STU model provide a good opportunity to 
comment on the interplay between symplectic deformations and Dirac charge quantization conditions induced by dyonic states.
When we include in the theory states with mutually non-local electric and magnetic charges such as black holes, the quantization condition on these charges breaks the duality symmetries to a discrete subgroup: $\Gd\to\Gd(\bbZ)_{\Gamma}$, preserving a certain lattice ${\Gamma}$.%
\footnote{For the purposes of this schematic discussion we will ignore the issue of what combinations of charges on a certain lattice are actually realized as states of the theory.
There is clearly an interplay with the gauging, as that affects the equations of motion and hence the kinds of solitonic (black-hole) solutions present in the theory.}
Moreover, symplectic redefinitions of the vector fields that preserve the lattice should similarly form a discrete group $\Sp(2\nv,\bbZ)_{\Gamma}$.
The FI gaugings also assign electric and magnetic charges to the fundamental fermions with respect to a certain vector field.
Therefore, the FI term must also belong to ${\Gamma}$.
All allowed FI terms compatible with this lattice will therefore be characterized by a space of the schematic form
\begin{formula}
\label{mSTU Sred discrete}
\fS_{\rm red}({\Gamma})=\left({\rm ISp}(6,\bbZ)_{\Gamma}\ \backslash\ \Sp(8,\bbZ)_{\Gamma}\ /\ \SL(2,\bbZ)_{\Gamma}^3\right)\,/\,(\bbS_3\times\bbZ_2^\pP)~.
\end{formula}
The discrete terms might also be affected, depending on ${\Gamma}$.
It is tempting to extend this expression to the general case \eqref{Sred general}, requiring consistent inequivalent gaugings to be related by the intersection of $\fS_{\rm red}$ with $\Sp(2\nv,\bbZ)_{\Gamma}$.
Notice that since we are discussing deformations of the embedding tensor, the right quotient corresponds to duality identifications, rather than field redefinitions of the scalar fields.

Of course, in this discussion the choice of ${\Gamma}$ still has to be specified.
The space of deformations is discretized only if other physical requirements fix it, because otherwise we are always allowed to deform ${\Gamma}$ together with the gauging.
This is consistent with the comments in \cite{DallAgata:2014ita}.

\subsection{Non-Abelian gaugings}

If we gauge a diagonal $\SL(2,\bbR)_{\rm gauge}\subset \SL(2,\bbR)_2\times \SL(2,\bbR)_3$, the vectors decompose into the representations
\begin{formula}
`A_{\mu}^a{\alpha}{\dot{\alpha}} \in ({\bf2},{\bf2},{\bf2}) \to ({\bf2},{\bf3}+{\bf1})\ \text{of}\ \SL(2,\bbR)_1\times \SL(2,\bbR)_{\rm gauge}\ .
\end{formula}
We can take  any linear combination of the two $\bf3$ representations to obtain a consistent gauge connection, and the broken $\SL(2,\bbR)_1$ symmmetries are sufficient to make any such choice equivalent, including the value of the gauge coupling constant.
Therefore, the gauge connection of a diagonal $\SL(2,\bbR)_{\rm gauge}$ is unique.\footnote{Notice however that there are two embeddings of $\SL(2,\bbR)_{\rm gauge}$ in \Gd: they differ by the action of the outer automorphism of $\SL(2,\bbR)_3$, which is not a symmetry.}

If we now combine the non-Abelian gauging with a FI term, the \U(1) must be gauged by an $\SL(2,\bbR)_{\rm gauge}$ singlet, so that only some combination of the vectors in the $({\bf2},{\bf1})$ can be used.
They take the form $`A_{\mu}^a {\epsilon}^{{\alpha}\dot{\alpha}}$.
We can use $\SL(2,\bbR)_1$ as above to fix the non-Abelian gauge connection entirely, and we are still left with the possibility to use its residual axionic shift to force the FI term to be either entirely electric or entirely magnetic.
Finally, the automorphism exchanging $\SL(2,\bbR)_2$ with $\SL(2,\bbR)_3$ can be used to flip the sign of the FI term.
We are therefore left with a discrete deformation corresponding to choosing the FI term entirely electric or entirely magnetic.
Moreover, the magnitude of the FI term with respect to the non-Abelian gauge coupling constant is a non-trivial, continuous parameter.

\section{Comments and conclusion}\label{sec6}

In this paper we have developed a general framework to classify four-dimensional gauged supergravities that share the same gauge group, but differ in the symplectic embedding of the gauge connection.
This framework is constructive in the sense that not only it provides all the necessary ingredients to build these supergravities explicitly, but it also determines what dualities and field redefinitions make certain models equivalent to each other.
All computations are group-theoretical in nature and it should be stressed that finite group elements play a central role in the construction, so that an analysis merely based on branchings of relevant Lie algebra representations is not a viable approach.

Using these tools we have investigated several examples of deformations of gauged supergravities with different amounts of supersymmetry.
We have focused our examples on reductive gauge groups, where it is easy to identify the discrete components of the relevant subgroups of \SPEM that appear in \eqref{Sred general}.
There is no obstruction to performing the very same computations for non-reductive \Gg, as was exemplified in \cite{DallAgata:2014ita} for maximal supergravity.

Studying the twin $\cN=6$ and $\cN=2$ truncations of the \SO(8) gauged maximal supergravities, we have found that despite the ${\omega}$-deformation is trivial in the $\cN=6$ theory (as also noted in  \cite{Borghese:2014gfa}), it is not in the $\cN=2$ gauged supergravity due to the presence of an extra Fayet--Iliopoulos coupling.
Moreover, the magnitude of this coupling with respect to the non-Abelian interactions can be changed, resulting in new $\cN=2$ models that cannot be uplifted to $\cN=8$.
These observations give rise to some interesting questions.
The two theories as obtained from $\cN=8$ share the same field content and couplings in the bosonic sector, which can therefore be lifted to the field content and equations of motion of type IIA supergravity on $\CP^3$ for any value of ${\omega}$, up to dualization of the singlet vector.
In particular, the vacua of these models are ${\omega}$-independent.
When fermions are included, however, things are subtler:
when ${\omega}=0$ the fermionic states captured by the $\cN=2$ theory are non-perturbative from the point of view of massless type IIA on $\CP^3$.
They have a perturbative interpretation in 11d supergravity, of course, being related to the two gravitini that restore maximal supersymmetry.
It becomes therefore tempting to ask whether the $\cN=2$ model for ${\omega}\neq0$ is including some non-perturbative fermionic states descending from some modification of type IIA on $\CP^3$.
This could be related to the generalized-geometric construction of \cite{Lee:2015xga}, where a generalized parallelization \cite{Lee:2014mla} satisfying the Leibniz algebra associated with the ${\omega}$ deformed \SO(8) gauging is constructed, that relies on a four-torus fibration over $\CP^3$.
Truncating to an \SO(6) sub-frame relevant for the truncation of type IIA supergravity, the dependence on the extra coordinates can be removed.
It would be very interesting to study how the $\cN=2$ theories constructed here can be related to such a construction, especially since only fermion fields should be expected to see the extended fibration.
If a consistent holographic dual to the ${\omega}$-deformation exists, the current findings suggest that it should be associated to a deformation of the sector of monopole operators of the ABJM theory responsible for its $\cN=8$ supersymmetry enhancement at levels $k=1,2$ \cite{Aharony:2008ug,Bashkirov:2010kz}.

While the rescaling of the $g_{\rm FI}/g_{\SO(6)}$ ratio in the same class of $\cN=2$ models breaks compatibility with an $\cN=8$ uplift, it does not necessarily break liftability to eleven-dimensional supergravity.
Contrary to ${\omega}$, this deformation could be associated with a vev of some \SO(6) singlet mode whose dynamics are truncated when reducing to four dimensions.
One natural guess would be the size of the Hopf circle of $S^1\hookrightarrow S^7\to\CP^3$, but it is worth noticing that \cite{Duff:1997qz} find three scalar \SO(6) singlets in the second massive level of the spectrum of eleven dimensional supergravity compactified on the Hopf circle.%
\footnote{There are no \SO(6) singlet scalar fields in the \SO(8) maximal supergravity spectrum: taking the gravitini to transform in the ${\bf8}_{\rm v}$ irrep, the relevant decomposition is ${\bf8}_{\rm v}\to2\times{\bf1}+{\bf6}$ and the scalar fields sit in the ${{\bf35}_{\rm c}+{\bf35}_{\rm s}\to 2\times{\bf15}+{\bf10}+\bar{\bf10}}$.}
Studying the scalar potential with both the $g_{\rm FI}/g_{\SO(6)}$ and ${\omega}$ deformations is a natural next step to understand the physics of these models.

In our classification of the gaugings of the STU model, we have found that no non-trivial ${\omega}$-like deformations are present for FI gaugings.
The one inherited from maximal supergravity can be removed by field redefinitions and electric-magnetic dualizations of the three ungauged vectors of the model.
This also means that asymptotically AdS black hole solutions of the STU model can be lifted not only to the standard \SO(8) theory, but also to the deformed ones, once the appropriate charge quantization conditions are imposed.
It would be interesting to find whether other supersymmetric, asymptotically AdS black holes can be found whose properties depend on ${\omega}$.
This would require to find such solutions in other supersymmetric truncations of the maximal theory where ${\omega}$ is non-trivial.

Since the trivial ${\omega}$ parameter of the STU model is only evident in the fermion couplings, we have also seen that it can be used to generate non-supersymmetric bosonic solutions from supersymmetric ones.
This trick of exploiting trivial deformations of an embedding tensor associated with broken compact symmetries of the theory could be straightforwardly applied to any other bosonic field configurations, including more general black holes in the STU or other models.

Finally, the large set of half-maximal gauged supergravities that we have constructed certainly deserve to be studied further.
A first step is surely to look for vacua in these theories.
The uplift of certain deformations to geometric and non-geometric backgrounds of string theory would also be extremely interesting:
it seems natural to expect that certain models would enjoy a description similar to certain gauged half-maximal supergravities in seven dimensions \cite{Dibitetto:2012rk,Lee:2015xga}.
The possibility to lift \SO(N) gauge groups (and other real forms)  geometrically on coset spaces as mentioned in \cite{Baguet:2015iou} would also be relevant.
We hope to come back to several of these questions in the near future.

\section*{Acknowledgements}
I thank Gianguido Dall'Agata, Bernard de Wit, Adolfo Guarino, Mario Trigiante and Daniel Waldram for discussions.
I thank Daniel Butter and Gianguido Dall'Agata for comments on a draft version of this manuscript.
This work was supported by the ERC Advanced Grant no. 246974, ``Supersymmetry: a window to non-perturbative physics''.

\bigskip

\providecommand{\href}[2]{#2}\begingroup\raggedright\endgroup

\end{document}